\documentclass[journal]{IEEEtran}

\usepackage{graphicx}
\usepackage{cite,amsmath,amsfonts,graphicx}
\usepackage{bm}
\usepackage{amsbsy}
\usepackage{color}
\usepackage{soul}

\begin{document}

\title{A Distributed Event-Triggered Control Strategy for DC Microgrids Based on Publish-Subscribe Model Over Industrial Wireless Sensor Networks}

\author{Seyed~Amir~Alavi,~\IEEEmembership{Graduate Student Member,~IEEE,}
            Kamyar~Mehran,~\IEEEmembership{Member,~IEEE,}
            Yang~Hao,~\IEEEmembership{Fellow,~IEEE,}
            Ardavan~Rahimian,~\IEEEmembership{Member,~IEEE,}
            Hamed~Mirsaeedi,~\IEEEmembership{Student Member,~IEEE,}
       	    and Vahid~Vahidinasab,~\IEEEmembership{Senior~Member,~IEEE}
       	    
\thanks{Manuscript received December 5, 2017; revised March 5, 2018, April 25, 2018, and June 30, 2018; accepted July 6, 2018.}
\thanks{\textsuperscript{\textcopyright} 2018 IEEE.  Personal use of this material is permitted.  Permission from IEEE must be obtained for all other uses, in any current or future media, including reprinting/republishing this material for advertising or promotional purposes, creating new collective works, for resale or redistribution to servers or lists, or reuse of any copyrighted component of this work in other works.}
\thanks{S. A. Alavi, K. Mehran, Y. Hao, and A. Rahimian are with the School of Electronic Engineering and Computer Science, Queen Mary University of London, London E1 4NS, U.K. (e-mail: k.mehran@qmul.ac.uk).}
\thanks{H. Mirsaeedi is with the School of Electrical and Computer Engineering, University of Tehran, Tehran, Iran.}
\thanks{V. Vahidinasab is with the Department of Electrical Engineering, Abbaspour School of Engineering, Shahid
Beheshti University, Tehran, Iran.}}

\markboth{IEEE Transactions on Smart Grid}%
{Shell \MakeLowercase{\textit{et al.}}: Bare Demo of IEEEtran.cls for IEEE Journals}

\maketitle

\begin{abstract}
This paper presents a complete design, analysis, and performance evaluation of a novel distributed event-triggered control and estimation strategy for DC microgrids. The primary objective of this work is to efficiently stabilize the grid voltage, and to further balance the energy level of the energy storage (ES) systems. The locally-installed distributed controllers are utilised to reduce the number of transmitted packets and battery usage of the installed sensors, based on a proposed event-triggered communication scheme. Also, to reduce the network traffic, an optimal observer is employed which utilizes a modified Kalman consensus filter (KCF) to estimate the state of the DC microgrid via the distributed sensors. Furthermore, in order to effectively provide an intelligent data exchange mechanism for the proposed event-triggered controller, the publish-subscribe communication model is employed to setup a distributed control infrastructure in industrial wireless sensor networks (WSNs). The performance of the proposed control and estimation strategy is validated via the simulations of a DC microgrid composed of renewable energy sources (RESs). The results confirm the appropriateness of the implemented strategy for the optimal utilization of the advanced industrial network architectures in the smart grids.
\end{abstract}


\IEEEpeerreviewmaketitle

\section{Introduction}

\IEEEPARstart{I}n the future modern smart grids, a direct current (DC) microgrid is becoming a natural substitution for the traditional power grid, mainly due to the ease of integration of the RESs. The DC microgrids have found the ever-increasing importance for the efficient realization of a number of crucial applications in the electric power industry, as they differentiate themselves from the AC grid counterparts in having the non-zero-crossing current and reactive power \cite{Zubieta2016a, Li2016a, Madduri2016}. However, controlling of the DC microgrid subsystems poses a considerable challenge, e.g., grid voltage stabilization is still a major problem due to the limitations in the voltage compensation techniques, when compared to the conventional AC power system \cite{Gungor2011,Justo2013,Planas2015}.

The microgrid control strategies are categorized as centralized, decentralized, and distributed \cite{Parhizi2015, Samad2017, Zohaib2018}. In the centralized architecture, all the data should be centrally processed where there is a single point of failure for the entire power system, which potentially reduces the reliability of the operation of the microgrid, as well as increasing the computational complexities of the measured data processing \cite{Olivares2014, Brandao2017, Minchala-Avila2016}. In the decentralized system architecture, a set of local controllers are distributed over the DC microgrid, e.g., the load controllers, distributed generation (DG) controllers, and converter controllers. The control objectives are achieved without any direct communication between the controllers, which is specifically useful in cases when the direct communication is not feasible or costly to establish. However, there are disadvantages to this method, such as the voltage offset and inadequate response time to load variations, which can both lead to the voltage and frequency instabilities in the DC microgrid. One of the most famous decentralized control methods is the droop control, where a number of investigations are conducted to overcome the deficiencies of the distributed control methods \cite{Iravani2016, Abessi2016, Vandoorn2012}.

The main advantages of a distributed control architecture include the seamless real-time operation, no single point of failure, reduction in the computational and communication complexities, and distribution of the tasks among the local controllers in the microgrid \cite{Morstyn2016, Wang2015}. The quality of power primarily relies on the fast real-time communication between the distributed controllers. Different approaches are proposed in order to effectively minimize the communication time and complexity. In \cite{Morstyn2016}, it is suggested that only one controller can communicate with its neighbor controllers rather than all the controllers. However, to satisfy the overall control objectives, the limitations of the main protocol candidates for the network should be appropriately considered, hence, the separation of concerns (SoCs) would not be held anymore \cite{Moayedi2016}.

The networked control systems (NCSs) are proved to be an essential framework for the implementation of a distributed control architecture in a number of applications, such as the power systems \cite{Singh2015}, industrial process control \cite{Wen2012}, power substation automation \cite{Montazeri2013}, aircraft control \cite{Wang2017}, and autonomous vehicles \cite{Herrera2011}. In an NCS, controllers are programmed on digital embedded platforms, and are connected to sensor/actuator nodes via a shared communication link. This offers a flexible structure where remote devices can be added, removed, and located with a minimum wiring and maintenance cost. With the advent of the Internet of things (IoT), the NCSs are moving towards wireless operation, where a channel may be shared among thousands of sensing nodes in the microgrid. Therefore, considering the fact that the channel bandwidth is limited, the NCS objectives should include reduction of the network traffic, as well as increasing the battery life of the sensors, while guaranteeing the overall performance. Hence, to fulfill such objectives and to address the issues regarding implementation of the intended control scheme, an event-triggered control and estimation strategy is further proposed in this work. The main objective in the event-triggered control is to efficiently reduce the communication packets generated in the sensor-controller-actuator loop \cite{Dormido2013}. Exchanging information can only be executed among the components, when an event is generated to satisfy the control performance. Furthermore, different datasets can be employed for the event-generation, e.g., the plant output signals \cite{Abdelrahim2016, Khashooei2017}, or state feedback signals \cite{Postoyan2015, Peng2013, Premaratne2013}.

Several works are already reported regarding the application of event-based control for microgrids. In \cite{Singh2017a}, a multi-agent control system is proposed which uses the reinforcement learning to estimate the state of other agents using the event-based message passing, with the objective of the load frequency control (LFC) in AC grids. Moreover, a similar approach is followed in \cite{Dong2017} using the adaptive dynamic programming (ADP) for LFC of AC microgrids. The authors in \cite{Zhou2017} have developed a power sharing control strategy among the hybrid microgrids using the event-based distributed consensus. Communication time delays and reactive power support are considered in the mentioned work. The problem of economic dispatch has been addressed in \cite{Li2016b} using the event-based communication and fast gradient distributed optimization among different microgrids, forming a resilient grid. The power sharing problem is addressed in \cite{Fan2017} to reduce the burden over the network. This idea has been extended in  \cite{Pullaguram2018, Sahoo2017} only for DC microgrids but without considering the communication models.

The main challenge of integrating the event-generation in an event-triggered control loop is the stabilization of the overall system. Several approaches are reported in the available literature, based on the use of the Lyapunov theorem for the stability analysis. In \cite{Tabuada2007}, to guarantee the global asymptotic stability, the problem of scheduling the event-based control tasks in the embedded processors is addressed. In \cite{Anta2010}, the event-triggered strategies to control the discrete-time systems are proposed, along with the results for the self-triggered control strategy. In \cite{Tabuada2007, Alavi2016}, the discrete-time systems are studied using a model-based Lyapunov triggering mechanism, in which the values of the Lyapunov functions are calculated at each sampling instant to maintain the system in the stable region. On the other hand, an event-triggered control system is considered as a hybrid system where the robustness analysis is rather a difficult task. In order to overcome the stabilization and robustness problems, we propose an approach by dividing the control problem into two different tasks. First, a distributed event-triggered optimal observer using the KCF \cite{Li2016} is designed to estimate the state of the DC microgrid via the sensing devices. Then, the distributed state feedback controller regulates the output voltage of the DC-DC converters to stabilize the average voltage of the microgrid to a set value.

\begin{figure}[!t]
\centering
\includegraphics[width=3.6in,height=3in]{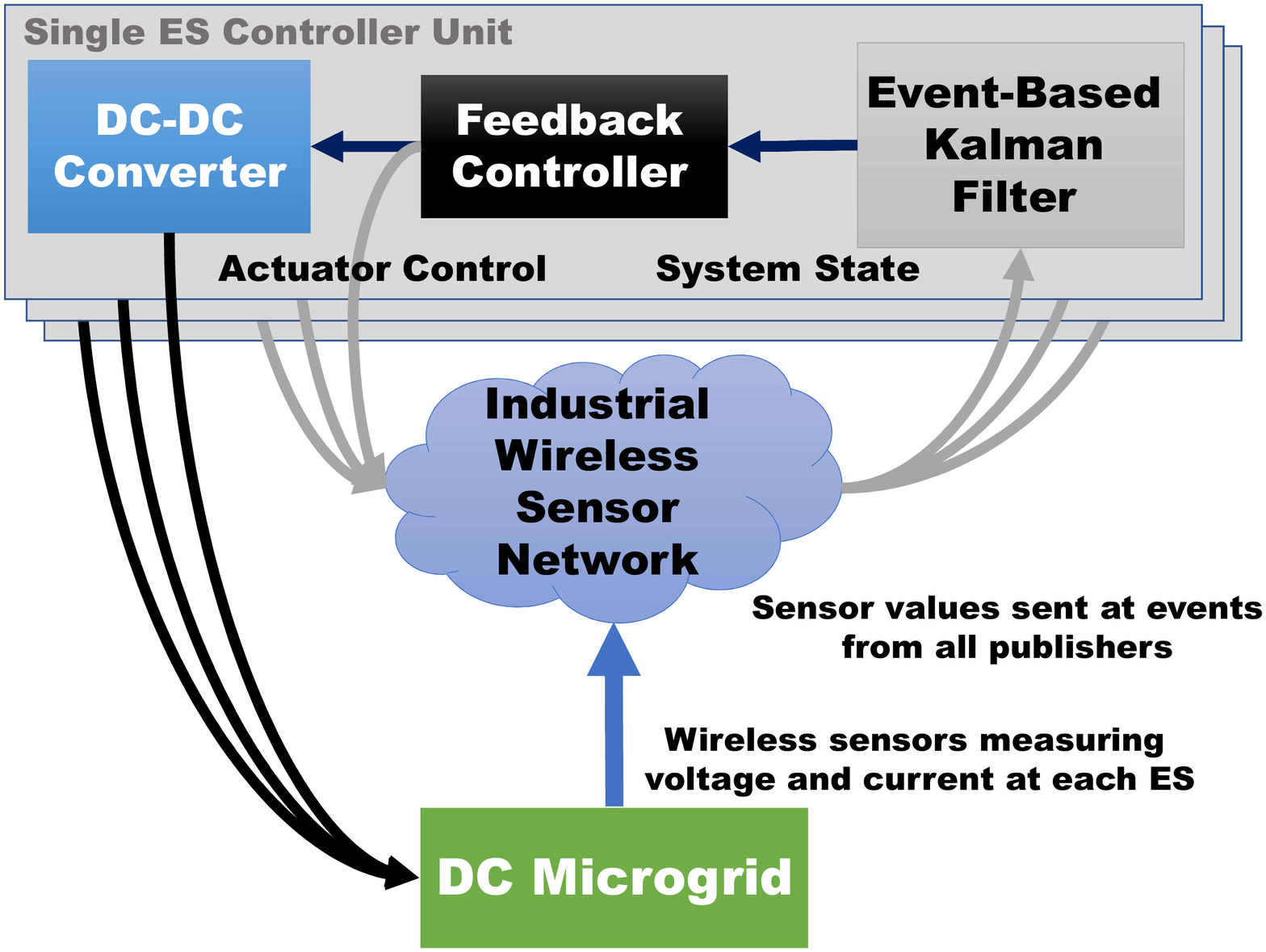}
\caption{Block diagram of the proposed distributed event-triggered control and estimation strategy for the DC microgrids.}
\label{escontrollerlbl}
\end{figure}

\figurename ~\ref{escontrollerlbl} presents the diagram of the distributed controller. A state estimator is proposed to reduce the number of data transmissions with relatively small degradation in the estimation performance. The send on delta (SoD) event-generation condition (i.e., $\delta$) is used in which the sensor data is transmitted only if its values go beyond the $\delta$ value. A case study of 10-bus microgrid is used to validate the proposed scheme, where the control objectives are chosen as the voltage stability and power sharing within the ES systems. Furthermore, multiple distributed ES systems replace a central ES system, to increase the reliability, power quality, and sensors' battery life, and to reduce the losses during the operation.

The contributions of this work can be summarized as:
\begin{enumerate}[\IEEEsetlabelwidth{12)}]
\item Regulating the voltage of the DC microgrid using a novel distributed control strategy, in order to effectively control the output voltages of the DC-DC converters connected to the ES systems. Also, the controller is fulfilling two objectives, i.e., balancing the energy level of the storages together with the voltage regulation.
\item Proposing an SoD-based Kalman filter, as a state estimator, for feedback control of the DC-DC converter, to balance the energy level by the distributed controller, along with the voltage regulation in the microgrid. The developed filter receives the real-time sensor data from the WSN, where the event-triggering function is based on the SoD sampling. The energy cost at each sensor node is also analyzed and compared to the traditional digital control system with the time-triggered sampling functionality. It is shown the network traffic is significantly reduced, due to the deployed procedure.
\item Utilizing the publish-subscribe model for the appropriate implementation of the event-based control strategy. It is shown that the model is seamlessly suitable for the event-based coordination of the distributed controllers.
\end{enumerate}

The rest of this paper is organized as follows. Section \ref{dcmicrogrid} describes the components of the DC microgrid, in which a microgrid model is developed based on the proposed distributed control and estimation strategy. A number of communication models have been thoroughly analyzed in terms of the event-based communication, and the publish-subscribe communication model is further proposed for the event-based sampling scenarios in Section \ref{pubsub}. In Section \ref{mgcontroller}, the proposed distributed controller design is discussed, and the structure of the Kalman filter as a state estimator is described, where a modification is suggested for the filter in order to adapt to the SoD event-based sampling. The stability and steady-state analyses are provided in Section \ref{globaldynamic}. In Section \ref{impl}, a case study is given to validate the performance of the controller using the simulations of a 10-bus DC microgrid. The paper is concluded in Section \ref{conc}.

\section{DC Microgrid Model Analysis}\label{dcmicrogrid}

\begin{figure}[!t]
\centering
\includegraphics[width=3.5in,height=3in]{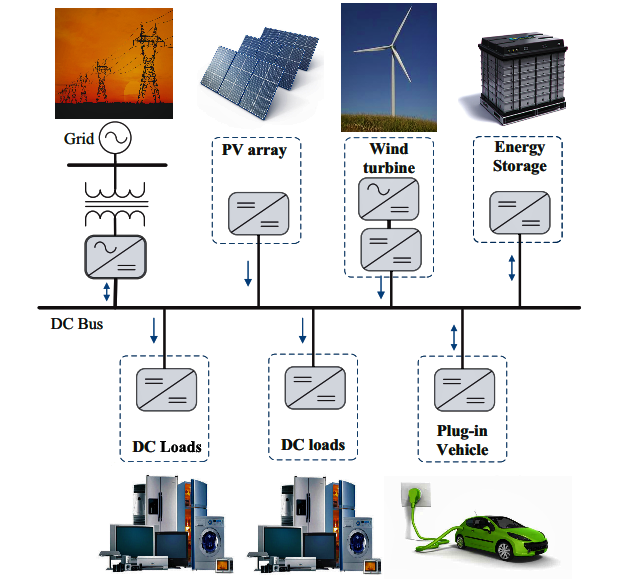}
\caption{Schematic of a DC microgrid with the main constituent components.}
\label{fig_sim}
\end{figure}

A DC microgrid essentially consists of four main components, including the DG, ES systems, power converters (DC-DC or DC-AC), and loads. \figurename ~\ref{fig_sim} indicates the common configuration of a microgrid, along with its constituent components.

The main objective of the ES systems is to compensate for the fast voltage dynamics caused by the load fluctuations in the DC microgrid. They are utilized to stabilize the voltage, and to increase the power quality. The widely used ES devices are electrochemical batteries, super capacitors, and flywheels which are easily deployable in a microgrid due to their natural DC output. A DC microgrid has two main operating modes: islanded mode and connected mode. The distributed secondary controller proposed in this work can operate in the both modes. A maximum power point tracking (MPPT) scheme is used to ensure that maximum power is absorbed from the intermittent distributed energy sources (DERs) such as photovoltaics. Here, the DGs and DC-DC converters in the main grid are modeled as the current sources, in which the variable injected current is related to their output power. In this way, the DC microgrid only gets connected to the main grid if enough power is not available from the installed DGs for load balance.

\section{Communication Models for Microgrids} \label{pubsub}
There are several communication models that can be integrated into a control system. Each model has both advantages and disadvantages, and the system designer has to effectively decide which model to employ for the implementation of the control system. In this work, several communication models are studied in detail and are compared with each other. In the following, a comparative analysis has been given for a number of the intended communication models, for the purpose of the appropriate employment in the DC microgrids.

\begin{enumerate}[\IEEEsetlabelwidth{12)}]
\item \textbf{Request/Response:} This communication model is one of the most commonly known models. It consists of a client that requests a service from a server, as shown in \figurename ~\ref{reqresmodel} (a). It is a useful model for the client-server or master-slave architectures \cite{Rodriguez-Dominguez2012}. However, a drawback of this model is the inequality of participants, which is apparent in the network topology. This makes it difficult for the bidirectional communication scenario, in which both the parties request information from each other, especially if firewalls are present. Consequently, either events, event-subscriptions, or security is difficult to manage, and require additional services and substantial resources if firewalls are used in the network.

\begin{figure}[!t]
\centering
\includegraphics[width=3.5in,height=3in]{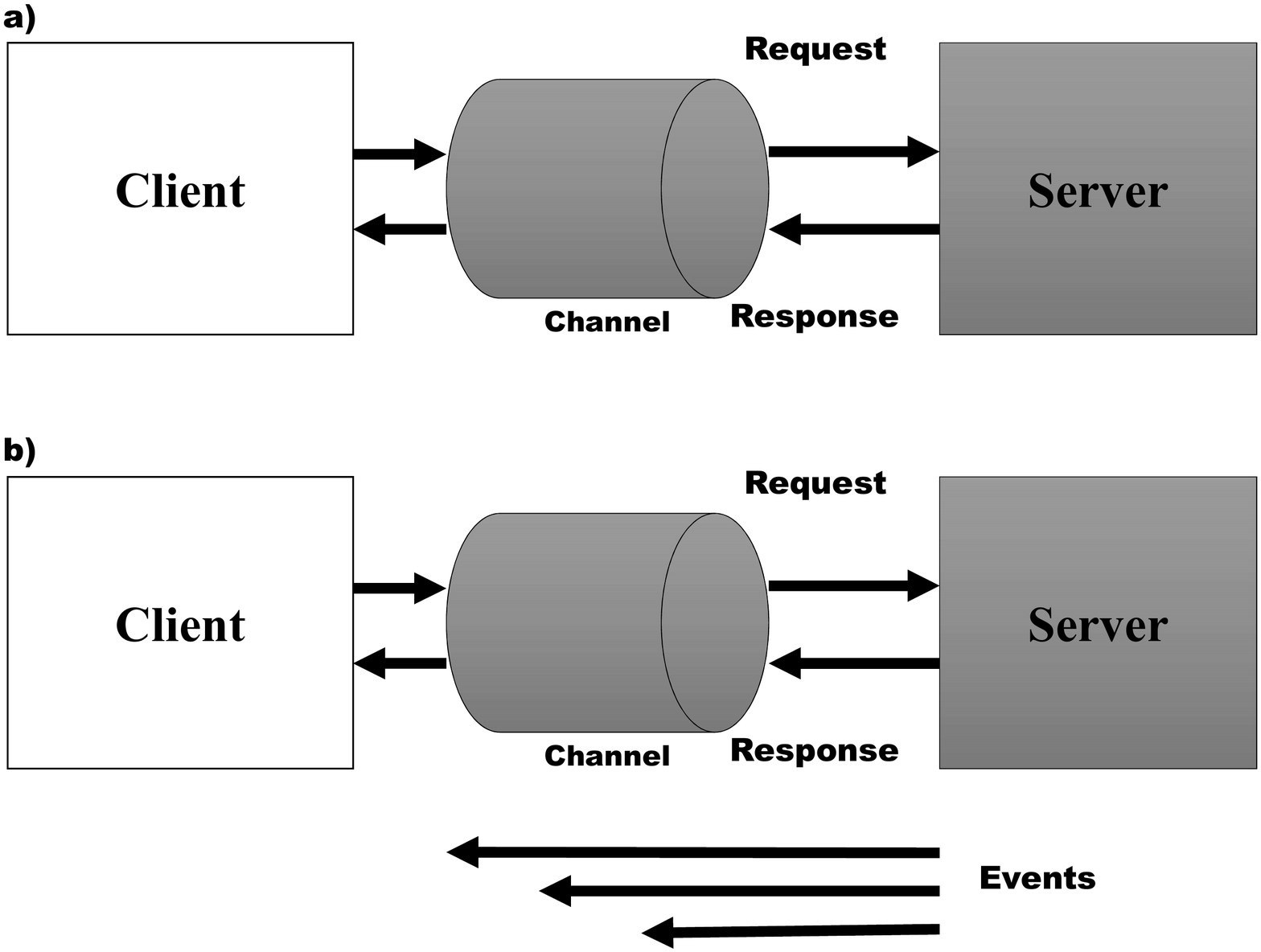}
\caption{The communication models for the bidirectional communication scenario: (a) request/response model; (b) event-subscription model.}
\label{reqresmodel}
\end{figure}

\item \textbf{Event-Subscription:}
This communication model allows a client to subscribe to events of a given type from a server. The server then informs the client each time the event is triggered, without having to constantly poll the server, as in \figurename ~\ref{reqresmodel} (b). Advanced event-subscription mechanisms can include client-­specific requirements of when events are desired and under what conditions. The benefits of using this communication model are that half of the messages are not needed over time, and the latency of updates is kept to a minimum. The problem with this model is that it is not designed for the multiparty communication scenario. This can be further solved using the publish-subscribe model.

\item \textbf{Multicasting:}
The previous models are primarily considered for the communication purposes between two entities. However, a more efficient model is required in cases when the same information has to be sent to multiple entities at the same time. Here, a sender sends one message through an intermediary (i.e., a broker or a router) which then distributes it to multiple recipients that have all requested participation in the communication. This model saves the bandwidth because the sender does not have to send individual messages to all the parties by itself. Also, the sender does not even have to know who the recipients are.

Although one can use this model in order to save the bandwidth, it is often used as a means of overcoming the restrictions in the chosen protocol, and its support of the event-subscription model, as well. In addition, multicasting is inherently difficult to secure, and it is more efficient in terms of the bandwidth only if the recipients actually use most of the transmitted values. In the case where frequent multicasting for decreasing the latency in the network is desired but not possible, the multicasting model might result in the increase rather than decrease in the required bandwidth \cite{Sakurai2012}.

\item \textbf{Queues:}
The first-in, first-out queues, is a model that allows one or more entities to post the messages or tasks into a queue, and then lets one or more receivers receive the messages in an ordered fashion. The queues reside on an intermediary node or network to which all participants are connected. This model is an excellent tool for the load balancing purposes, where the collected tasks from multiple sources need to be distributed among the existing workers, perhaps having different performances. Queues can hardly be used for real-time communications in control systems, since the message should be saved at first, and then be processed at the controller via an intermediary node.
\end{enumerate}

\textbf{Publish/Subscribe:}
This communication model is an extension of the multicasting model, with the difference that messages transmitted are stored in the intermediary node. The messages, or a reference to the messages, are distributed to the corresponding subscribers, depending on the protocol. Also, only the latest message is stored, a given number of messages are stored, or all messages are stored in the intermediary, depending on the chosen protocol, as well as the settings of the intermediary \cite{Liu2009}. The difference between distributing the entire message and distributing only a reference to the message is important and affects the performance of the solution in terms of the consumed bandwidth. If the subscribers consume most of the messages, forwarding the messages themselves is more efficient, as in the case of multicasting. If, however, consumption occurs only on demand, then sending shorter references is more efficient because these messages are smaller and subscribers would use only a minority of them to fetch an actual message. In order to fetch a message in the latter case, a separate request/response action needs to be performed \cite{Flores2011}.

The behavior of each model has been analyzed from the control point of view. In this treatise, the publish-subscribe communication model is used for the practical implementation of the distributed event-based control strategy. In the publish-subscribe model, a node can act as a publisher, subscriber, or both simultaneously. The network roles can be dynamically changed to ensure a flexibility to reconfigure the directions of the data exchange. The main advantage of this model is that the data can be exchanged intelligently between the devices (i.e., the publishers send the data to the specific subscribers without having a subscription knowledge of each node). This keeps the setup process easier for the overall maintenance of the network, and enables the self-configuration of the devices, as one of the primary characteristics of the industrial ad-hoc networks. The process of selecting messages for the reception and processing is called filtering. The topic-based and content-based filtering are the two common forms of filtering used in new communication protocols introduced in the context of IoT. In the publish-subscribe network setup, a server manages the topics and contents, which is called a broker. The broker-free setup can be achieved with the distributed topics/contents suitable for the proposed distributed control structure \cite{Roffia2016}. The topic-based publish-subscribe communication model also enables the selective message distribution among a number of sources and sinks \cite{Xylomenos2012}. Messages are associated with the topics and are selectively routed to destinations with matching topic interests. Subscribers show their interest in receiving data with a given topic and data sources publish messages on the topics.

The main advantages of the publish/subscribe communication model compared with the aforementioned models can be summarized as:

\begin{itemize}

\item Adaptive role change in a dynamic environment from the publisher to the subscriber and vice versa.
\item Intelligent data exchange among the nodes without having a subscription knowledge of each node.
\item Automatic self-configuration of the nodes in the ad-hoc network without a central configurator which enables the plug and play operation of the microgrid.
\item Intrinsic discrete event transmission support which suits it as an ideal choice for the event-triggered control.
\end{itemize}

\begin{figure}[!t]
\centering
\includegraphics[width=3.5in,height=2.8in]{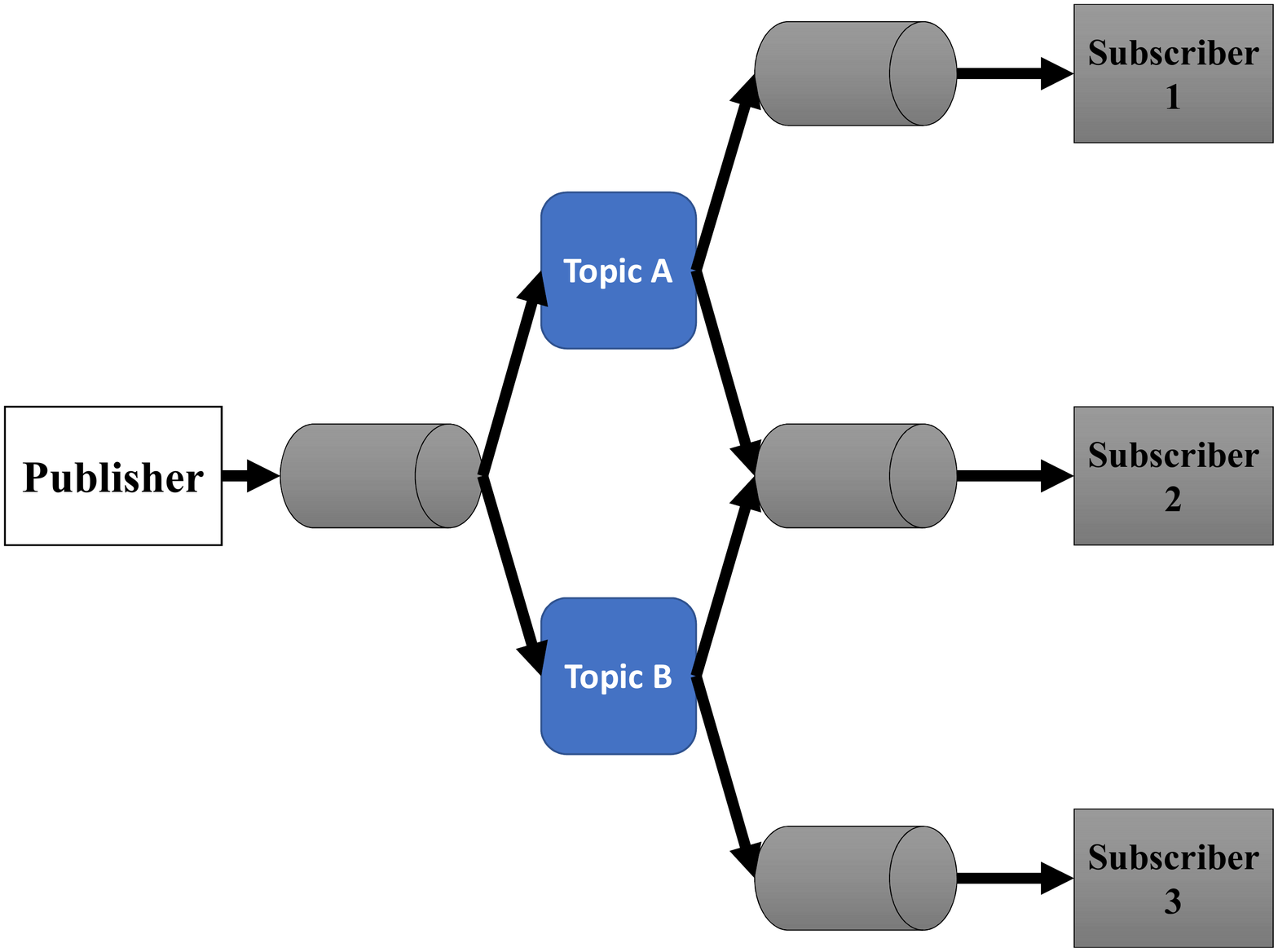}
\caption{Block diagram of the topic-based publish-subscribe model for the industrial distributed communication scenario.}
\label{pubsublbl}
\end{figure}

\begin{figure}[!t]
\centering
\includegraphics[width=3.5in,height=2.8in]{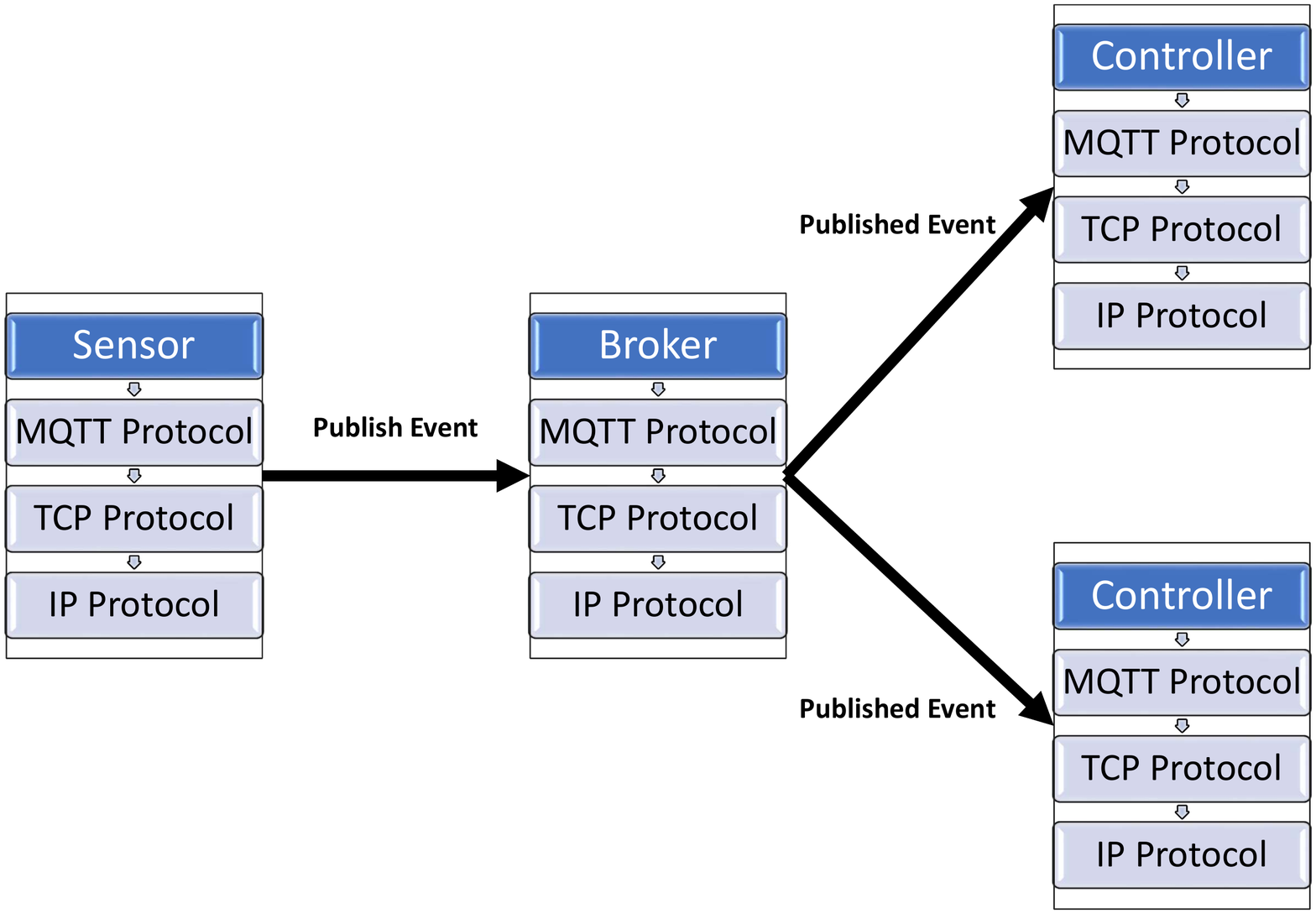}
\caption{The MQTT protocol stack for the event-based implementation.}
\label{mqtt_implementationlbl}
\end{figure}

\figurename ~\ref{pubsublbl} also presents the concept of the topic-based publish-subscribe communication protocol model. Multiple subscribers can listen for a predetermined topic, and also multiple publishers can publish new data to certain topics. The only drawback with this model is that when subscribers initially subscribe to a certain topic or content, their initial value remains undefined until the next publishing cycle. A number of communication protocols are proposed to tackle this issue, such as the message queue telemetry transport (MQTT) protocol, which uses the retained value in the broker-based structures. Consequently, when a subscriber connects to the broker, it will release the retained value of the most updated publish to the subscriber \cite{Alavi2018, Monajemi2017}. The protocol stack of MQTT is depicted in \figurename ~\ref{mqtt_implementationlbl}.

\begin{table*}[!t]
\renewcommand{\arraystretch}{1.3}
\caption{Comparison Between the Different Communication Models for the Distributed Event-Triggered Control of DC Microgrids.}
\centering
\begin{tabular}{|c|c|c|c|c|}
\hline
Communication Model & Support for Dynamic Environment & Smart Message Delivery & Plug \& Play Capability & Event Transmission Support\\
\hline
Request/Response & Weak & Weak & Not supporting & Not supporting\\
Event-Subscription & Weak & Medium & Not supporting & Strong\\
Multicasting & Strong & Medium & Medium & Medium\\
Queues & Medium & Medium& Not Supporting & Medium\\
\textbf{Publish/Subscribe} & \textbf{Strong} & \textbf{Strong} & \textbf{Strong} & \textbf{Strong}\\
\hline
\end{tabular}
\end{table*}

\section{Distributed Control Strategy Analysis}\label{mgcontroller}

A bidirectional DC-DC boost converter provides an interface between the ES system and the DC microgrid. The boost converter also acts as a bidirectional charger. A voltage-current (V-I) droop controller regulates the DC-DC converter output voltage with the reference voltage of $v^*_i$, which is calculated based on the microgrid voltage reference $v^{mg}$ and the locally measured output current $i_i$. In the proposed strategy, the droop control is improved by considering two extra control signals $u^{\overline{v}}_i$, $u^e_i$ in the current reference signal $v^{\mathrm{*}}$: 

\begin{equation}
v^{\mathrm{*}}_i\mathrm{=}v^{mg}\mathrm{-}F_ir_i\mathrm{(}i_i\mathrm{-}u^{\overline{v}}_i\mathrm{-}u^e_i\mathrm{)}
\end{equation}
where the voltage stabilization control signal $u^{\overline{v}}_i$ is defined to regulate the average microgrid bus voltage, and the power sharing control signal $u^e_i$ is proposed to balance the energy level between the ES systems and to maintain it through the load sharing. In a decentralized V-I droop control, the load is normally shared between the ES systems in inverse proportion to their virtual resistances $r_i$. The virtual resistance (i.e., $r_i$) is merely used for the voltage regulation at the converters of the ES systems, therefore it is lossless.

The DC microgrid is subject to high frequency harmonics due to the pulse width modulation (PWM) switching control scheme used for the converters. A low-pass filter with the cut-off frequency of ${\omega }^c_i$ should be used to reduce the harmonics, and to prevent the resulting instabilities in the grid.
\begin{equation}
F_i\mathrm{=}\frac{{\omega }^c_i}{{s\mathrm{+}\omega }^c_i}
\end{equation}
where ${\omega }^c_i$ is a time-variant parameter considering the switching control of the DC-DC converters. However, with an appropriate selection of an upper-bound value, it can be approximated to a time-invariant parameter.

The current regulation can be achieved in two stages \cite{Mayo-Maldonado2011}. At the first stage, a proportional-integral (PI) voltage controller $G^v_i$ is defined in (\ref{eq:1}) to set the converter current reference to regulate the output voltage of the ES system: 
\begin{equation} \label{eq:1}
i^{\mathrm{*}}_i\mathrm{=}G^v_i\left(v^{\mathrm{*}}_i\mathrm{-}v_i\right)\mathrm{,\ }\mathrm{\ }G^v_i\mathrm{=}p^{vp}_i\mathrm{+}\frac{p^{vi}_i}{s}
\end{equation}
At the second stage the current controller sets the duty cycle of the PWM switching to control the bipolar junction transistors (BJTs) of the converter, and to regulate the output current.

In order to balance the energy level, a PI controller $G^e_i$ is defined in (\ref{eq:25}) to set the energy level $e_i$ to the local estimate of the average energy level of the ES systems. Due to different capacities for the energy storages, per-unit energy level is used for the power balancing signal.
\begin{equation}\label{eq:25}
u^e_i\mathrm{=}G^e_i\left(e_i\mathrm{-}{\overline{e}}_i\right)\mathrm{,\ \ }G^e_i\mathrm{=}p^{ep}_i\mathrm{+}\frac{p^{ei}_i}{s}
\end{equation}

Another PI controller $G^{\overline{v}}_i$ is used for the voltage regulation of the microgrid, where the local estimate of the average bus voltage is regulated to the voltage reference of the microgrid:
\begin{equation}\label{eq:mg_voltage_control}
u^{\overline{v}}_i\mathrm{=}G^{\overline{v}}_i\left(v^{mg}\mathrm{-}{\overline{v}}_i\right)\mathrm{,\ }\ G^{\overline{v}}_i\mathrm{=}p^{\overline{v}p}_i\mathrm{+}\frac{p^{\overline{v}i}_i}{s}\mathrm{+}\frac{p^{\overline{v}ii}_i}{s^{\mathrm{2}}}
\end{equation}

In (\ref{eq:mg_voltage_control}), the double-integral is used to maintain the overall stability, and to eliminate the steady-state error. An average state estimator is designed for each ES system, using the local measurements and information from the neighboring ES systems. The estimator updates the local estimates of the average energy level and bus voltage of the ES system, then the controller tries to regulate the average estimates to the nominal values of the microgrid.

\subsection{Distributed Average Consensus Protocol}
Each ES system has an average state estimator that uses the local measurements and information from the neighboring ES systems to update the local estimates of the average ES system per-unit energy level $\overline{e}_i$, average microgrid bus voltage $\overline{v}_i$, and average ES system output current. The average state estimator implements a distributed average consensus protocol for tracking the dynamic signals from \cite{Spanos2005DynamicCF}.

The ES systems are connected by a sparse communication graph $\mathcal{G(V,E)}$ with the nodes $\mathcal{V}=(1,...,\mathcal{N)}$ and edges $\mathcal{E}$. Each graph node represents an ES system, and the graph edges represent communication links between them. $(i,j)\in \mathcal{E}$ if there is a link allowing information flow from node $i$ to node $j$. The neighbors of $i$ node are given by $\mathcal{N}_i$, where $j \in \mathcal{N}_i$ if $(j,i)\in \mathcal{E}$. The graph adjacency matrix is given by $A = [a_{ij}] \in R^{N \times N}$, where $a_{ij} > 0$ if $(j,i)\in \mathcal{E}$ and $a_{ij}=0$ otherwise.

For the $i$th ES system, let $x_i$ be a local state variable, and let $\overline{x}_i$ be a local estimate of the average value of that state for the ES systems. The $i$th ES system receives the average state estimates from its neighbors $j \in \mathcal{N}_i$, and its average state estimator implements the following distributed average consensus protocol:
\begin{equation}
\overline{x}_i=x_i+\int \sum\limits_{j \in \mathcal{N}_i}{a_{ij}(\overline{x}_j-\overline{x}_i)dt}
\end{equation}

Each node in the network has in-degree $d_i =\sum_{j=1}^{N} a_{ij}$ and out-degree $d_i^o = \sum_{j=1}^{N} a_{ji}$. Moreover, the graph is balanced if $d_i = d_i^o$ for all the nodes. The graph degree matrix is given by $\textbf{D} = diag\{d_i\}$ and the graph Laplacian matrix is also given by $\textbf{L}=\textbf{D}-\textbf{A}$. The global dynamics of the distributed average consensus protocol are given by:
\begin{equation}
\dot{\overline{\textbf{x}}} = \dot{\textbf{x}}-\textbf{L}\overline{\textbf{x}}
\end{equation}

Applying the Laplace transform yields the following transfer function matrix for the average consensus protocol \cite{Spanos2005DynamicCF}:
\begin{equation}
G^{avg}=\frac{\overline{\textbf{X}}}{\textbf{X}}=s(sI_N+\textbf{L})^{-1}
\end{equation}

$\overline{\textbf{X}}$ and $\textbf{X}$ are the Laplace transforms of $\overline{x}$ and $x$, respectively.

For a balanced communication graph with a spanning tree, the steady-state gain of the average consensus protocol is given by the averaging matrix \cite{Nasirian2015}:
\begin{equation}
\lim_{s\to 0} G^{avg}= Q \mathrm{,where}\ [Q]_{ij}=\frac{1}{N}
\end{equation}
The final value theorem shows that for a vector of step inputs, the elements of $\overline{\textbf{x}}(t)$ converge to the global average of the steady-state values $\textbf{x}^{ss}$:
\begin{equation}\label{eq:dafafasaxz}
\lim_{t\to \infty} \overline{\textbf{x}}(t)= \lim_{s\to 0} G^{avg}\lim_{t\to \infty} s\textbf{X}=Q\textbf{x}^{ss}=\left\langle \textbf{x}^{ss} \right\rangle \underline{1} 
\end{equation}

\subsection{Event-Based Kalman Filter Design}
Consider the following linear system which is the state space realization of distributed average consensus protocol transfer function in each controller:
\begin{IEEEeqnarray}{ll}
\dot{x}=Ax\left(t\right)+w\left(t\right)\IEEEnonumber \\ y(t)=Cx(t)+v(t)
\end{IEEEeqnarray}
where $x\mathrm{\ }\mathrm{\in}\mathrm{\ }{\mathrm{R}}^{\mathrm{n}}$ is the estimated state and $y\ \mathrm{\in}\mathrm{\ }{\mathrm{R}}^{\mathrm{p}}\mathrm{\ }$is the output measurement. The process noise $w\left(t\right)$ and measurement noise $v(t)$ are the uncorrelated, zero-mean white Gaussian random signals, fulfilling the following:
\begin{IEEEeqnarray}{lll}
E\left\{w(t) \; w(s)'\right\} = Q \; \delta (t-s)
\\
E\left\{v(t) \; v(s)'\right\} = R \; \delta (t-s)
\\
E\left\{w_i\left(t\right){v_j\left(s\right)}'\right\} =0,\;\;1 \leq i \leq n, \;\;1 \leq j \leq p
\end{IEEEeqnarray}
where $w_i$ and $v_j$ are the \textit{i}-th and \textit{j}-th elements of the $w$ and $v$, respectively. Also, $R$ is the measurement noise covariance, and $Q$ is the process noise covariance. It is assumed that the\textit{ i}-th sensor only transmits the data when the difference between the current sensor value and the previously transmitted value is greater than ${\delta }_i$.

The states are also estimated periodically with the period of $T$. For simplicity, it is assumed that there is no delay in the sensor data transmission. Using the SoD method, the estimator continuously with a period of $T$ demands the data from the sensors no matter the data becomes available. For example, if the last received $i$-th sensor value is $y_i$ at the time $t_{last,i}$, and there is no $i$-th sensor data received for ${t>t}_{last,i}$, then the estimator can estimate $y_i(t)$ as:
\begin{equation}
y_i\left(t_{last,i}\right)-{\delta }_i\le \ y_i\left(t\right)\le y_i\left(t_{last,i}\right)+{\delta }_i
\end{equation}

The last received \textit{i}-th sensor data is used to compute the output $y_{computed,i}$ even if there is no sensor data transmission:
\begin{equation}
\label{eq:2}
y_{computed,i}\left(t\right)=y_i\left(t_{last,i}\right)=C_ix\left(t\right)+v_i\left(t\right)+{\Delta }_i\left(t,t_{last,i}\right)
\end{equation}
where ${\mathrm{\Delta }}_i\left(t,t_{last,i}\right)\mathrm{=}y_i\left(t_{last,i}\right)\mathrm{-}y_i\left(t\right)$ and: 

\begin{equation}
\label{eq:3}
\left|{\Delta }_i\left(t,t_{last,i}\right)\right|\le {\delta }_i
\end{equation}

In (\ref{eq:2}), the measurement noise increases from $v_i\left(t\right)$ to $v_i\left(t\right)+{\Delta }_i\left(t,t_{last,i}\right)$. If ${\Delta }_i\left(t,t_{last,i}\right)$ is assumed to have the uniform distribution with (\ref{eq:3}), then the variance of ${\Delta }_i\left(t,t_{last,i}\right)$ is $\frac{{\delta }^{\mathrm{2}}_i}{\mathrm{3}}$, which is added to the \textit{measurement noise covariance} in standard Kalman filter R$\left(i,i\right)$ when (\ref{eq:2}) applies.

\textbf{Improved Kalman Measurement Update Algorithm:} An algorithm is proposed here to appropriately improve the \textit{measurement update} part of the standard Kalman filter algorithm, which is adapted to the SoD event-generation condition by increasing the measurement noise covariance $\overline{R}_k$:
 \begin{enumerate}
 \item  Initialization set
 \begin{IEEEeqnarray}{ll}
 \hat{x}^-(0),{P}^-_0 \IEEEnonumber \\ 
 y_{last}=C\hat{x}^-\left(0\right)
 \end{IEEEeqnarray}
 \item  \textbf{Measurement update
 \begin{equation}
 \overline{R}_k=R
 \end{equation}
 if \textit{i}-th measurement data are received
 \begin{equation}
 \hat{y}_{last,i}=y_i\left(kT\right)
 \end{equation}
else
\begin{equation}
\overline{R}_k\left(i,i\right)=\overline{R}_k\left(i,i\right)+\frac{{\delta }^2_i}{3}
\end{equation}
end if
\begin{IEEEeqnarray}{lll}
K_k={P}^-_kC'(C{P}^-_kC'+\overline{R}_k)^{-1}\IEEEnonumber \\ 
\hat{x}\left(kT\right)=\hat{x}^-\left(kT\right)+K_k(\hat{y}_{last}-C\hat{x}^-(kT))\IEEEnonumber\\
P_k{=(I-K_kC)P}^-_k
\end{IEEEeqnarray}
}
\item  Project ahead
\begin{IEEEeqnarray}{lll}
\hat{x}^-\left((k+1)T\right)=\exp{\left(AT\right)}\hat{x}\left(kT\right)\IEEEnonumber\\
{P}^-_{k+1}=\exp{\left(AT\right)} P_k\exp{\left(A'T\right)}+Q_d
\end{IEEEeqnarray}
 \end{enumerate}
where $Q_d$ is the process noise covariance for the discretized dynamic system; $y_{last}$ is defined as (\ref{eq:jhjf}):
\begin{equation}\label{eq:jhjf}
y_{last}=[y_{last,1},y_{last,2},\dots ,y_{last,p}]'
\end{equation}

The presented event-triggered Kalman filter has been developed to implement the distributed controller and estimator as an NCS. It should be noted that in the proposed event-triggered observer, convergence is obtained by using the Kalman optimal observer. However, choosing the lower values of ${\delta }_i$ would result in the considerable reduction in the convergence time \cite{Li2016}. The controllers only receive updates from their neighbor controllers which is reflected in the $\textbf{L}$ matrix of the transfer function that has been realized. Distributed average consensus is then achieved for each controller based on the number of neighbor controllers. Also, the higher the number of adjacent controllers are, the faster the estimator would converge.

\section{Global Dynamics and Stability Analysis}\label{globaldynamic}

\begin{figure}[!t]
\centering
\includegraphics[width=3.5in,height=2.5in]{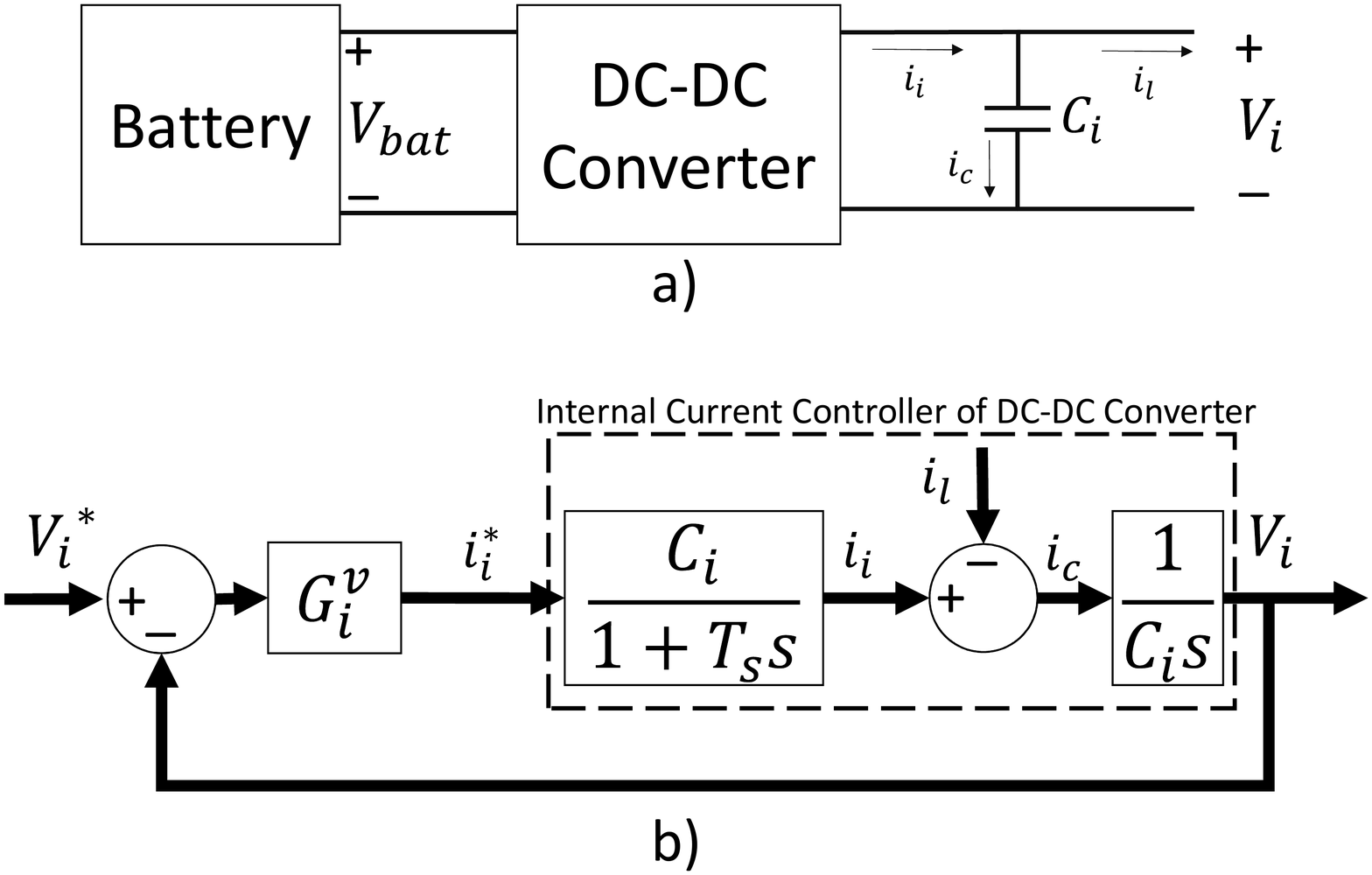}
\caption{Internal model of the ES system: (a) DC-DC converter circuit; (b) block diagram of the local converter controller.}

\label{figblockdiagram}
\end{figure}

\figurename ~\ref{figblockdiagram} presents the block digram of the feedback loop for each of the distributed ES controllers. The voltage regulation dynamics of the grid forms a multiple-input multiple-output (MIMO) linear system. If $V^{mg}$ is the Laplace transform of the voltage reference of the grid, the distributed control dynamics can be expressed as (\ref{eq:globaldynamic}):
\begin{equation}\label{eq:globaldynamic}
V^*=V^{mg}\underline{1}-Fr\left(I-G^{\overline{v}}\left(V^{mg}\underline{1}-\overline{V}\right)-G^e\left(E-\overline{E}\right)\right),\ \
\end{equation}
where
\begin{IEEEeqnarray}{lllll}
F=diag\left\{F_i\right\}\\r=diag\left\{r_i\right\}\\
G^{\overline{v}}=diag\left\{{G_i}^{\overline{v}}\right\}\\ G^e=diag\left\{G^e_i\right\}\\
\overline{V}=G^{avg}V\ and\ \overline{E}=G^{avg}E 
\end{IEEEeqnarray}

\begin{figure}[!t]
\centering
\includegraphics[width=3.5in,height=2.6in]{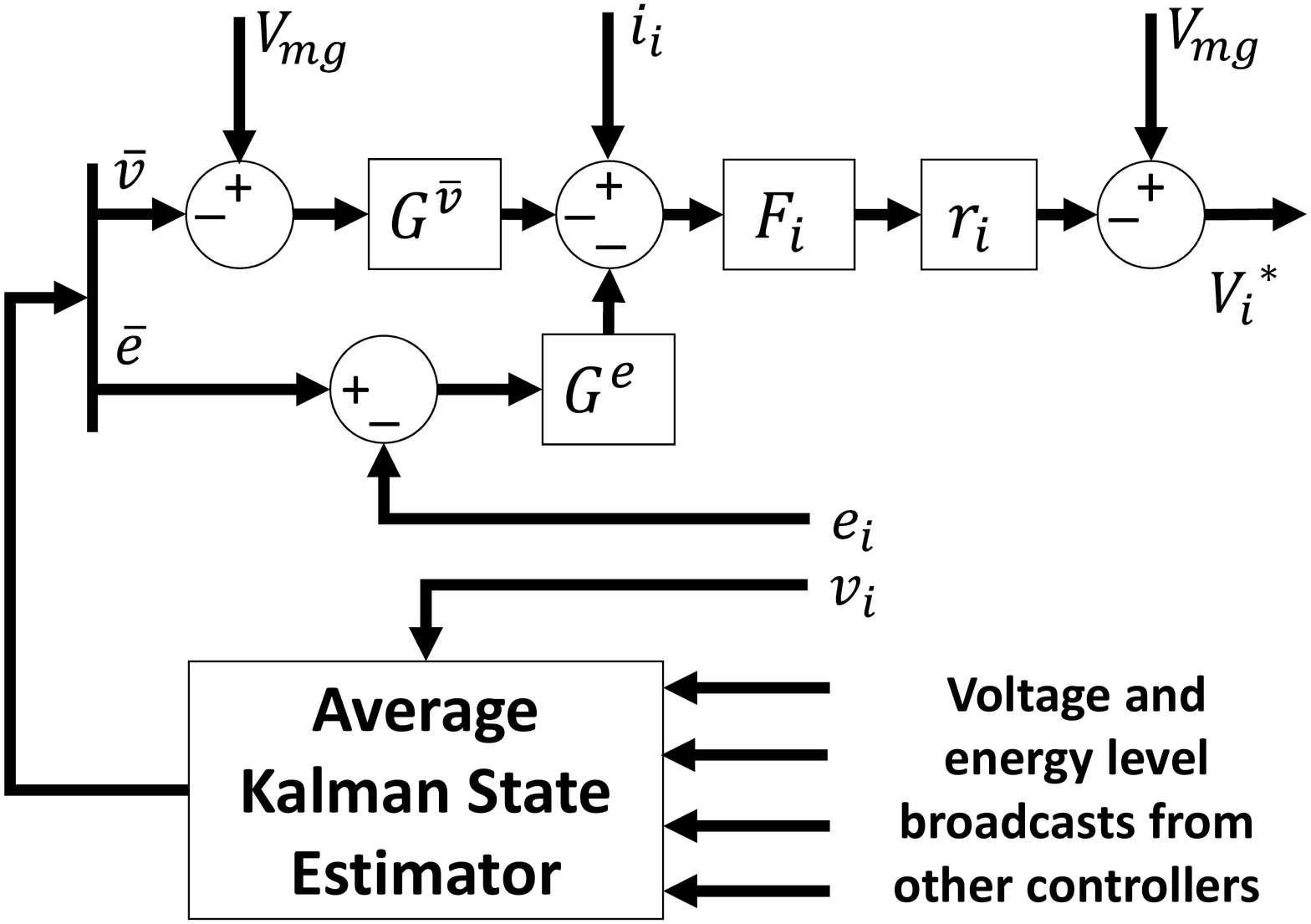}
\caption{Block diagram of the proposed distributed feedback controller.}
\label{twoloopcontroller}
\end{figure}

The grid-connected rectifier, the constant power loads, as well as the generation sources (i.e., operate under the MPPT algorithm) can act as the positive or negative current sources, while the ES systems act as the bus voltage regulation units in the DC microgrid. To formulate the bus voltage regulation dynamics, power sources can be modeled by a parallel current source and resistance. Modern DC-DC converters operate at a high switching frequency with one switching period delay (i.e., $T_s$) in the current control mode. In order to model the DC-DC converter, a control structure as shown in \figurename ~\ref{twoloopcontroller} is used, in which the bus voltage regulation dynamics is designed as an outer-loop between the output voltage of the ES system $v^{\mathrm{*}}_i$, and the local bus voltage $v_i$ \cite{Chen2012a}. Moreover, the transfer function for the internal loop is given by $H^{v_{ol}}_i$.

\begin{equation}\label{eq:sadsad}
H^{v_{cl}}_i=\frac{H^{v_{ol}}_i}{1+H^{v_{ol}}_i}\ ,\ \ H^{v_{ol}}_i=\frac{G^v_i}{sC_i(T_ss+1)}
\end{equation}
Therefore, the local bus voltage closed-loop transfer function of the DC microgrid is given by:
\begin{equation}
V\mathrm{=}H^{v_{cl}}V^{\mathrm{*}}\mathrm{,\ }{\ \ \ \ \ \ \ \ H}^{v_{cl}}\mathrm{=}diag\{H^{v_{cl}}_i\}
\end{equation}
The output currents of the ES system can be obtained from multiplying the bus voltages with the bus admittance matrix, constructed based on the line and load impedances:
\begin{equation}
I=YV
\end{equation}
A first order model is used for the battery per-unit energy level charging and discharging:
\begin{equation}
{\dot{e}}_i=-\frac{v_ii_i}{e^{max}_i}
\end{equation}
where $e^{max}_i$ is the maximum energy capacity of the \textit{i}-th ES system. The global energy level dynamics is modeled as (\ref{eq:xcvcvx}):
\begin{equation}\label{eq:xcvcvx}
E=MYV,\ \ M=diag\{-\frac{v^{mg}}{e^{max}_is}\}
\end{equation}
The global closed-loop voltage regulation dynamics can be described by the multiple output linear system as (\ref{eq:hghg}):
\begin{IEEEeqnarray}{ll}\label{eq:hghg}
V=[{\left(H^{v_{cl}}\right)}^{-1}+FrY+FrG^{\overline{v}}G^{avg}\IEEEnonumber \\ 
-FrG^e\left(I_N-G^{avg}\right)MY]^{-1}V^{mg}\left(I_N+FrG^{\overline{v}}\right)\underline{1}
\end{IEEEeqnarray}

In the above strategy, it is assumed the local distributed controllers can exchange data with the other controllers in a continuous mode. It should also be noted that this assumption is not feasible in the cases involving the NCSs. An event-based Kalman filter is proposed to overcome this problem. Using this filter, the distributed control system can be realized with the SoD event triggering condition.
\\

\subsection{Stability and Steady-State Analysis}
Assuming $v^{mg}$ be the control reference voltage. In this case, input to the global closed-loop voltage dynamics is given by:
\begin{equation}\label{eq:kjljad}
V^{mg}=\frac{v^{mg}}{s}
\end{equation}

The steady-state DC microgrid bus voltages are obtained by applying the final value theorem to (\ref{eq:kjljad}):

\begin{IEEEeqnarray}{ll}\label{eq:uouoi}
v^{ss}= \lim_{s\to 0} sV 
\IEEEnonumber \\ 
=\lim_{s\to 0} [s^2\left(H^{v_{cl}}\right)^{-1}+s^2FrY +s^2FrG^{\overline{v}}G^{avg}
\IEEEnonumber \\  
-s^2FrG^e\left(I_N-G^{avg}\right)MY]^{-1}s^2v^{mg}\left(I_N+FrG^{\overline{v}}\right)\underline{1}
\IEEEnonumber \\  
\end{IEEEeqnarray}

The steady-state bus voltages can be reached to based on the following limits:

\begin{IEEEeqnarray}{ll}\label{eq:conditions}
\lim_{s\to 0} s^2G^{\overline{v}}=G^{\overline{v}ii}, \mathrm{where}\ G^{\overline{v}ii}=\mathrm{diag}\{k_i^{\overline{v}ii}\}
\IEEEnonumber \\  
\lim_{s\to 0} sG^{e}=G^{ei}, \mathrm{where}\ G^{ei}=\mathrm{diag}\{k_i^{ei}\}
\IEEEnonumber \\  
\lim_{s\to 0} sM=M_{0}, \mathrm{where}\ M_{0}=\mathrm{diag}\{-\frac{v^{mg}}{e_i^{max}}\}
\IEEEnonumber \\ 
\lim_{s\to 0} G^{avg}=Q, \lim_{s\to 0} Y=Y_{0},\ \lim_{s\to 0} F=I_{N},
\IEEEnonumber \\ 
\mathrm{and} \lim_{s\to 0} {\left(H^{v_{cl}}\right)}^{-1}=I_{N}
\IEEEnonumber \\ 
\end{IEEEeqnarray}
therefore
\begin{IEEEeqnarray}{ll}
v^{ss}=\left[r\left(G^{\overline{v}ii}Q-G^{ei}(I_N-Q)M_0Y_0\right)\right]^{-1}v^{mg}(rG^{\overline{v}ii})\underline{1}
\IEEEnonumber \\ 
\end{IEEEeqnarray}
which yields
\begin{IEEEeqnarray}{ll}\label{eq:xdasdsa}
\left[(G^{ei})^{-1}G^{\overline{v}ii}Q-(I_N-Q)M_0Y_0\right]v^{ss}=v^{mg}(G^{ei})^{-1}G^{\overline{v}ii})\underline{1}
\IEEEnonumber \\ 
\end{IEEEeqnarray}

Furthermore, as shown in (\ref{eq:conditions}), without the double-integral gain of the voltage controller, the steady-state response would be dominated by the energy balancing control signal; to verify that the average steady-state voltage is equal to the reference voltage of the microgrid, each side of (\ref{eq:xdasdsa}) is multiplied by the averaging matrix $Q$. Since the column sums of $(I_N-Q)$ are equal to zero, $Q(I_N-Q)=0_{N \times N}$. Following (\ref{eq:dafafasaxz}) yields:

\begin{IEEEeqnarray}{c}
Q\left((G^{ei})^{-1}G^{\overline{v}ii}Qv^{ss}\right)=v^{mg}Q\left((G^{ei})^{-1}G^{\overline{v}ii}\underline{1}\right)
\IEEEnonumber \\ 
\left\langle v^{ss}\right\rangle\left\langle(G^{ei})^{-1}G^{\overline{v}ii}\underline{1}\right\rangle\underline{1} = v^{mg}\left\langle (G^{ei})^{-1}G^{\overline{v}ii}\underline{1}\right\rangle\underline{1}
\IEEEnonumber \\
\left\langle v^{ss}\right\rangle = v^{mg}
\end{IEEEeqnarray}

\section{Results and Discussion}\label{impl}

The performance evaluation of the proposed controller is thoroughly presented in this section through a case study of a 10-bus microgrid. As also depicted in \figurename ~\ref{escontrollerlbl}, each distributed controller receives the events from neighbor ES sensors. The deployed sensors measure the bus voltages and currents.

It has been discussed in detail that the network traffic, and the battery energy usage of sensor nodes in a WSN, would be reduced significantly if an event-triggered strategy is used. The event-triggered control stops the unnecessary data exchange in a shared medium. Once an event is generated, the data must be sent to the controller as fast as possible, in order to prevent the deviation of system behavior from the stable margin.

In the proposed distributed control, there are two different variables that are evaluated in the event-generation condition. First is the bus voltage in which the distributed controller is installed, and second is the ES system per-unit energy level. The conditions of the SoD event-generation for these variables are independent, therefore two thresholds are evaluated in each controller. Each event is then matched with its corresponding topic in the publish-subscribe model; e.g., in the presented case study of the 10-bus system, each bus controller publishes data in two topics related to that bus. Since the network is assumed to be connected, each distributed controller subscribes to the topics of the other neighbors. This is shown in Table \ref{tab:bustopic}. The MATLAB \& Simulink software is employed for the simulation of the DC microgrid and distributed control strategy. Also, the Simscape toolbox is used to simulate the electrical distribution system of the DC microgrid.
\begin{table}[!t]
\renewcommand{\arraystretch}{1.3}
\caption{Distributed Controllers and Their Corresponding Topics.}
\label{tab:bustopic}
\centering
\begin{tabular}{|c|c|c|}
\hline
Bus Controller & Voltage SoD Event Topic & Energy SoD Event Topic\\
\hline
Bus 1 & voltageBus1 & energyBus1\\
\hline
Bus 2 & voltageBus2 & energyBus2\\
\hline
Bus 3 & voltageBus3 & energyBus3\\
\hline
Bus 4 & voltageBus4 & energyBus4\\
\hline
Bus 5 & voltageBus5 & energyBus5\\
\hline
Bus 6 & voltageBus6 & energyBus6\\
\hline
Bus 7 & voltageBus7 & energyBus7\\
\hline
Bus 8 & voltageBus8 & energyBus8\\
\hline
Bus 9 & voltageBus9 & energyBus9\\
\hline
Bus 10 & voltageBus10 & energyBus10\\
\hline
\end{tabular}
\end{table}

\subsection{DC Microgrid Configuration}
\begin{figure}[!t]
\centering
\includegraphics[width=3.5in,height=4.9in]{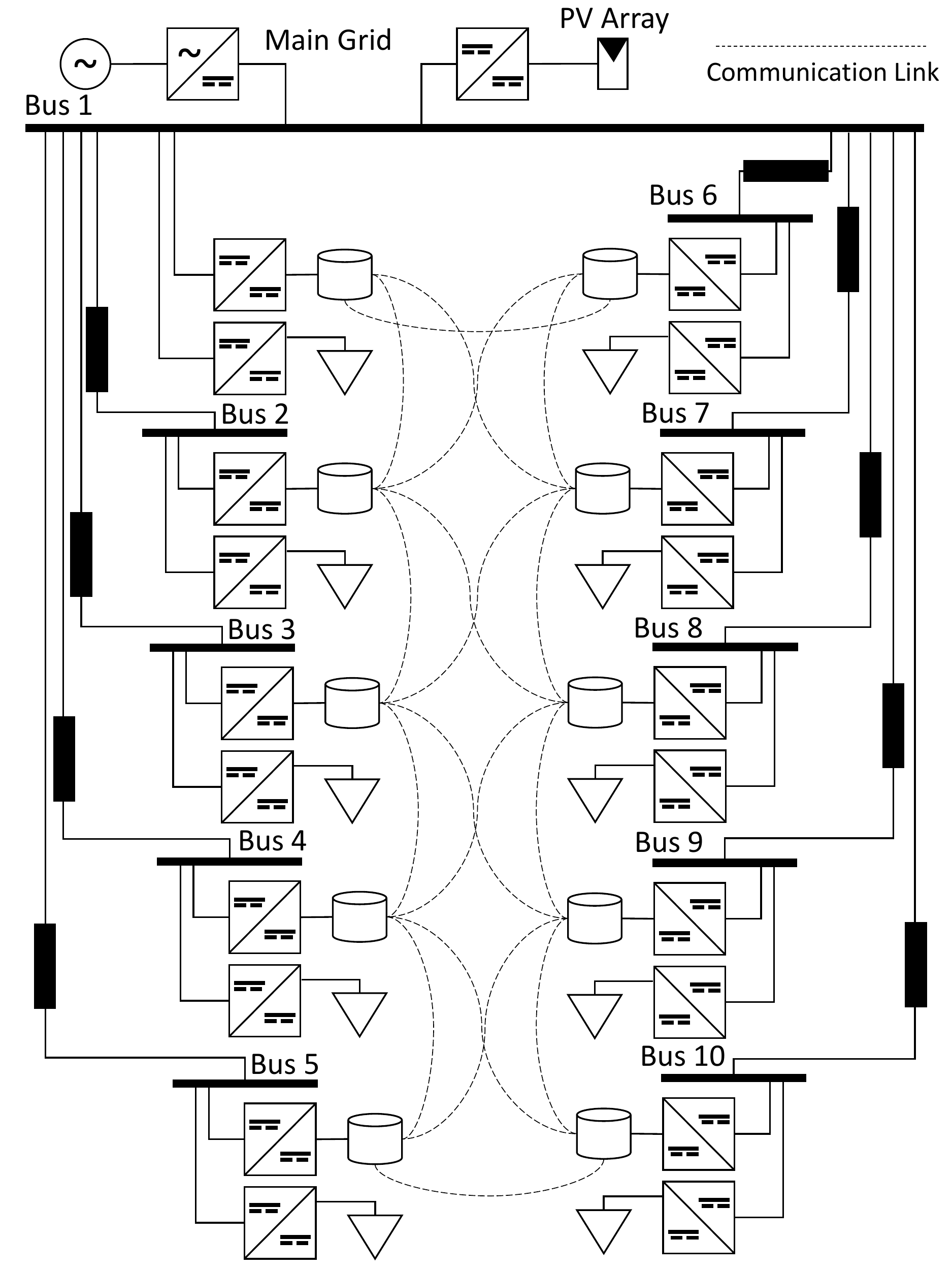}
\caption{Proposed case study of the 10-bus DC microgrid with the ES systems.}
\label{casegridlbl}
\end{figure}

The microgrid used for the case studies is shown in \figurename ~\ref{casegridlbl}. The presented DC microgrid incorporates a 10-bus distribution system with the PV generation and 10 battery ES systems.

At bus 1, a 150 kW rated rectifier provides main connection of the microgrid. Bus 1 also includes 500 $m^2$ PV generation operated with the MPPT algorithm, rated for 80 kW. Based on the analysis of conventional wiring configurations of the DC microgrids shown in \cite{Tanaka2012} for data centers, $50m\times24mm$ cables are selected to connect the load buses to bus 1. The buses 1 to 7 have 25 kWh lithium-ion batteries, while the buses 8 to 10 ES systems have 12.5 kWh lithium-ion batteries. The battery ES systems are connected by a sparse communication network to support the proposed distributed event-triggered control. The communication links between the ES systems are bidirectional, meeting the requirements of the distributed control strategy for a balanced communication network. Based on the ETSI EN 300 132-3-1 telecommunications DC distribution standard for data centers, the voltage limits are defined as 380 V$\pm 5\%$  \cite{Becker2011}.

\begin{figure}[!t]
\centering
\includegraphics[width=2.5in,height=1.8in]{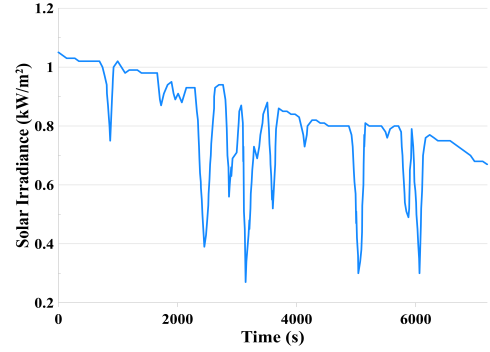}
\caption{Data of the PV solar irradiance used in this case study.}
\label{pvdynamics}
\end{figure}

For the case study, 15 kW constant power loads are installed at buses 1 to 5, and 5 kW constant power loads are installed at buses 6 to 10, hence the total load of the microgrid would be 100 kW. The battery ES systems begin with values around half of their energy levels, and are chosen randomly. The bus 1 PV generation with MPPT was simulated based on the modeling approach from \cite{Villalva2009}, using the 1 min resolution irradiance and temperature data for 2 p.m. to 4 p.m. on June 1, 2014 from the NREL Solar Radiation Research Laboratory (SRRL): Baseline Measurement System (BMS), in Colorado. Moreover, the used irradiance data is shown in \figurename ~\ref{pvdynamics}. The simulation parameters are also provided in Table \ref{tab:paramstbl}. The graph adjacency matrix $A$ elements $a_{ij}$ are chosen as "1" if there is a connection and "0" if there is no communication link between the buses. The DC load and battery parameters for each bus are given in Table \ref{tab:loadbattbl}.  The parameters of the proposed event-based Kalman filter are also provided in Table \ref{tab:kalmanparamstbl} for the proposed control strategy.

\begin{table}[!t]
\renewcommand{\arraystretch}{1.3}
\caption{Values of the Loads and Batteries at Each Bus.}
\label{tab:loadbattbl}
\centering
\begin{tabular}{|c|c|c|}
\hline
Bus & Load Power & Battery Capacity (380 V)\\
\hline
Bus 1 & 15 kW & 25 kWh\\
\hline
Bus 2 & 15 kW & 25 kWh\\
\hline
Bus 3 & 15 kW & 25 kWh\\
\hline
Bus 4 & 15 kW & 25 kWh\\
\hline
Bus 5 & 15 kW &25 kWh\\
\hline
Bus 6 & 5 kW &25 kWh\\
\hline
Bus 7 & 5 kW &25 kWh\\
\hline
Bus 8 & 5 kW &12.5 kWh\\
\hline
Bus 9 & 5 kW &12.5 kWh\\
\hline
Bus 10 & 5 kW &12.5 kWh\\
\hline
\end{tabular}
\end{table}

\subsection{Simulation Scenario}
The simulation scenario is divided into four sections to represent the different modes of operation of the proposed control strategy. The scenario is simulated for both with and without the proposed event-triggered estimation, in order to compare the results of both the implementations. Moreover, in another simulation, the performance of the proposed estimator is tested by adding 100 ms delay in the event transmission. The simulation time is set at 120 minutes (i.e., 7,200 seconds).

\subsubsection{Islanded Operation With Load Switching, 0 to 10 min}
The DC microgrid begins in the islanded mode. The start load is 60\% at all the employed buses. After 5 minutes, the loads are switched at all buses to their 100\% nominal values. As shown in \figurename ~\ref{simenergy}, the energy level of the ES systems at the buses are increased, as the total power from the PV exceeds the total load of the microgrid. After the load switching, storages starting to discharge the energy, and the voltage is stabilized around 380 V with the zero error, as shown in \figurename ~\ref{simvoltage} (C).

\subsubsection{Grid Connected Operation  With Rectifier Providing Load Balancing, 10 to 40 min}
At min 10, a grid connection is made with the rectifier in the load balancing mode. Moreover, the ES systems use their 30 kW power capacity to reach a balanced per-unit energy level, as shown in \figurename ~\ref{simenergy}. A per-unit energy level of 0.45 is reached by the ES systems; 11.25 kWh for the 25 kWh ES systems at buses 1 to 7, and 5.62 kWh for the 12.5 kWh ES systems at buses 8 to 10. The voltage controllers limit the bus voltages of the DC microgrid between 377.3 and 381.6 V (i.e., 1\% error), and ensure the average bus voltage remains at the voltage reference of the microgrid, as shown in \figurename ~\ref{simenergy} (B). As desired, the average ES system per-unit energy level remains constant around the operating point.

\subsubsection{Grid Connected Operation With Main Grid Providing ES Charging, 40 to 80 min}
\begin{figure*}[tb]
\includegraphics[width=\textwidth,height=13cm]{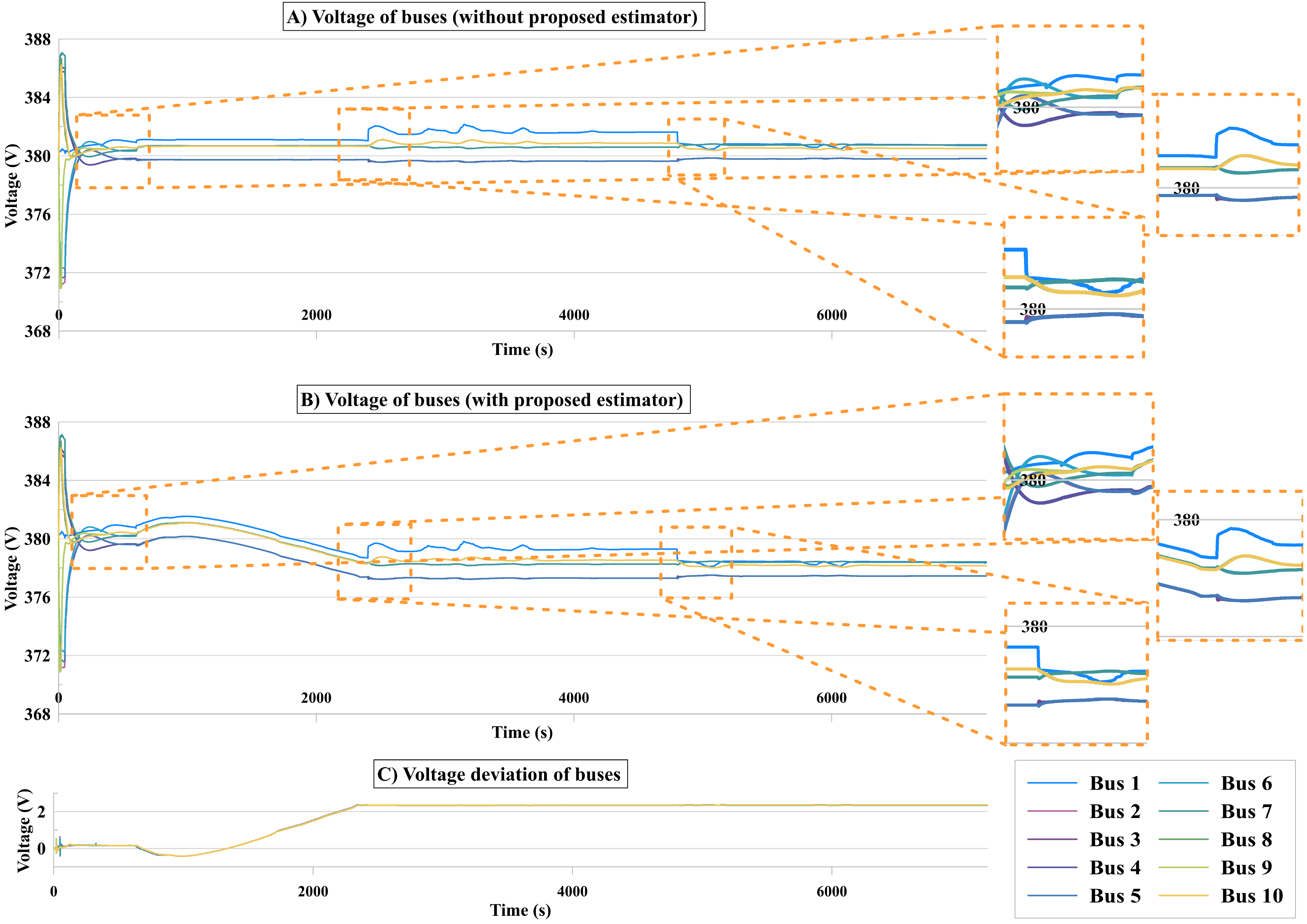}
\caption{Simulation results of the bus voltages for the case study of the 10-bus DC microgrid: A) without the proposed estimator; B) with the proposed estimator; C) error between the two approaches.}
\label{simvoltage}
\end{figure*}

At min 40, the rectifier operating mode is changed from the load balancing to the ES charging mode, and the injected power increases. The rectifier uses its maximum power capacity of 150 kW to raise the average ES system per-unit energy level to the value of 0.62, as shown in \figurename ~\ref{simenergy} (B). The per-unit energy balancing is maintained between the ES systems. The 25 kWh ES systems are charged at a common rate and the 12.5 kWh ES systems are charged at half of this rate. As the ES systems are charged, they adjust their output powers to balance the variable PV generation, and to further regulate the average DC microgrid voltage within 0.05 V of the reference of 380 V, as shown in \figurename ~\ref{simvoltage} (B).

\subsubsection{Islanded Operation With Sudden Main Grid Disconnection, 80 to 120 min}
At min 80, the grid connected rectifier is suddenly disconnected, initiating the islanded operation. The sudden power imbalance causes the bus voltages of the DC microgrid to fall, with a minimum level of 377.4 V reached. The ES systems react to the fall in the voltage by increasing the corresponding output powers, restoring the microgrid load balance and returning the average bus voltage to the reference with less than 1\% (i.e., 2 V) steady-state deviation.

\begin{table}[!t]
\renewcommand{\arraystretch}{1.3}
\caption{Parameters of the Case Study and Controller.}
\label{tab:paramstbl}
\centering
\begin{tabular}{|c|c|c|c|c|c|}
\hline
$R_{dc}$ & 36 m$\Omega$ & Voltage & 380 V & $ p_i^{\bar{v}p}$ & 500\\
\hline
$L_{dc}$ & 7 $\mu H$ &  $p_i^{vi}$ & 10 & $ p_i^{\bar{v}i}$ & 10\\
\hline
$r$ & 0.2533 & $w_i^c$ & 100 rad/s & $ p_i^{\bar{v}ii} $ & 0.1\\
\hline
$p_i^{vp}$ & 10 & $ p_i^{ep}$ & 5000 & $ p_i^{ei}$ & 50\\
\hline
\end{tabular}
\end{table}

\begin{table}[!t]
\renewcommand{\arraystretch}{1.3}
\caption{Parameters of the Event-Triggered Kalman Filter.}
\label{tab:kalmanparamstbl}
\centering
\begin{tabular}{|c|c|}
\hline
$\delta_i  (Voltage)$ &  0.1 V \\
\hline
$\delta_i (Energy)$ &  0.01 p.u. \\
\hline
$Q$ & 0 \\
\hline
$R$ & 1 \\
\hline
$T$ & 1 Second \\
\hline
\end{tabular}
\end{table}

\begin{figure*}[tb]
\includegraphics[width=\textwidth,height=13cm]{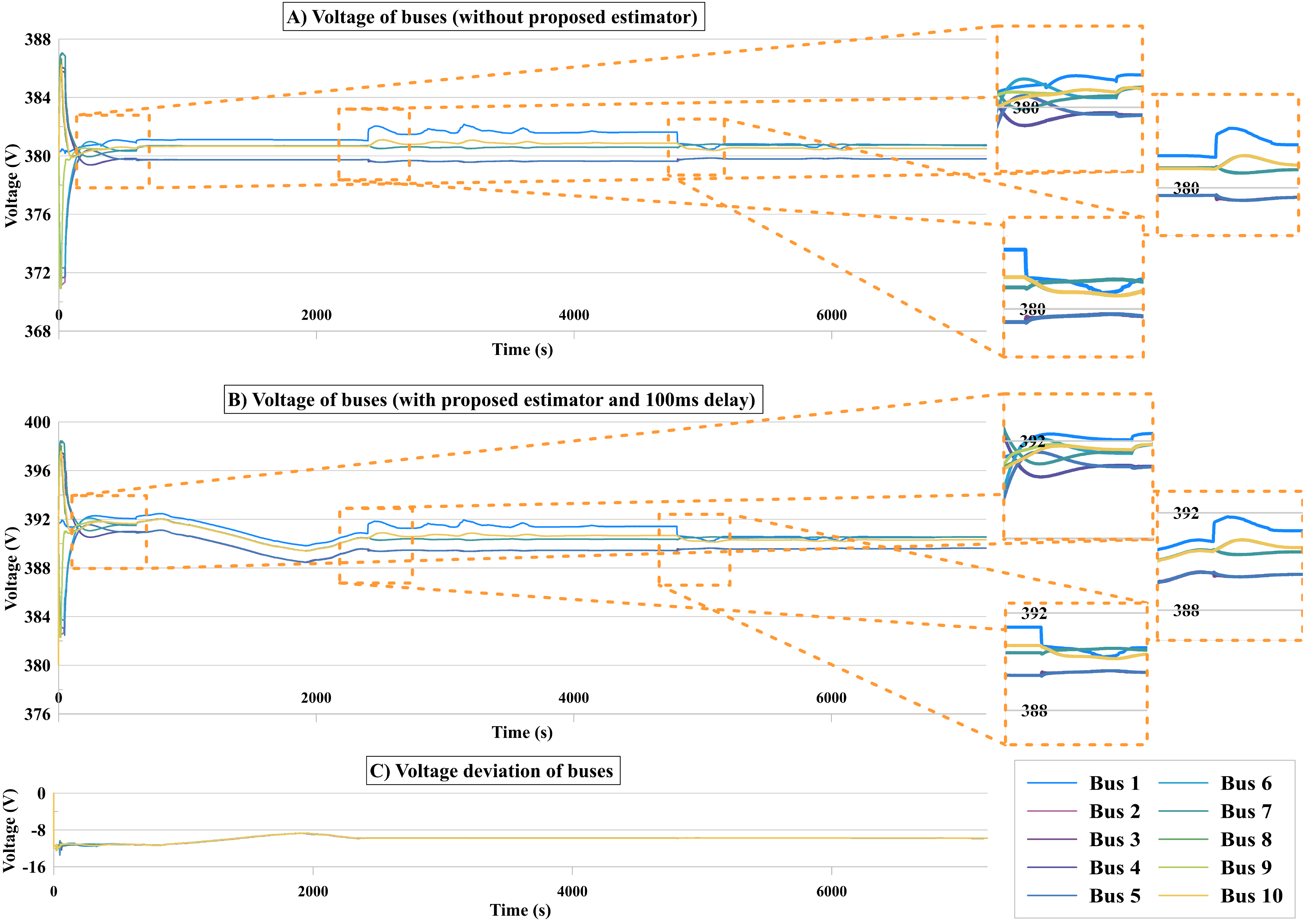}
\caption{Simulation results of the bus voltages for the case study of the 10-bus DC microgrid: A) without the proposed estimator; B) with the proposed estimator (100 ms delay); C) error between the two approaches.}
\label{simvoltagedelay}
\end{figure*}

\begin{figure*}[tb]
\includegraphics[width=\textwidth,height=13cm]{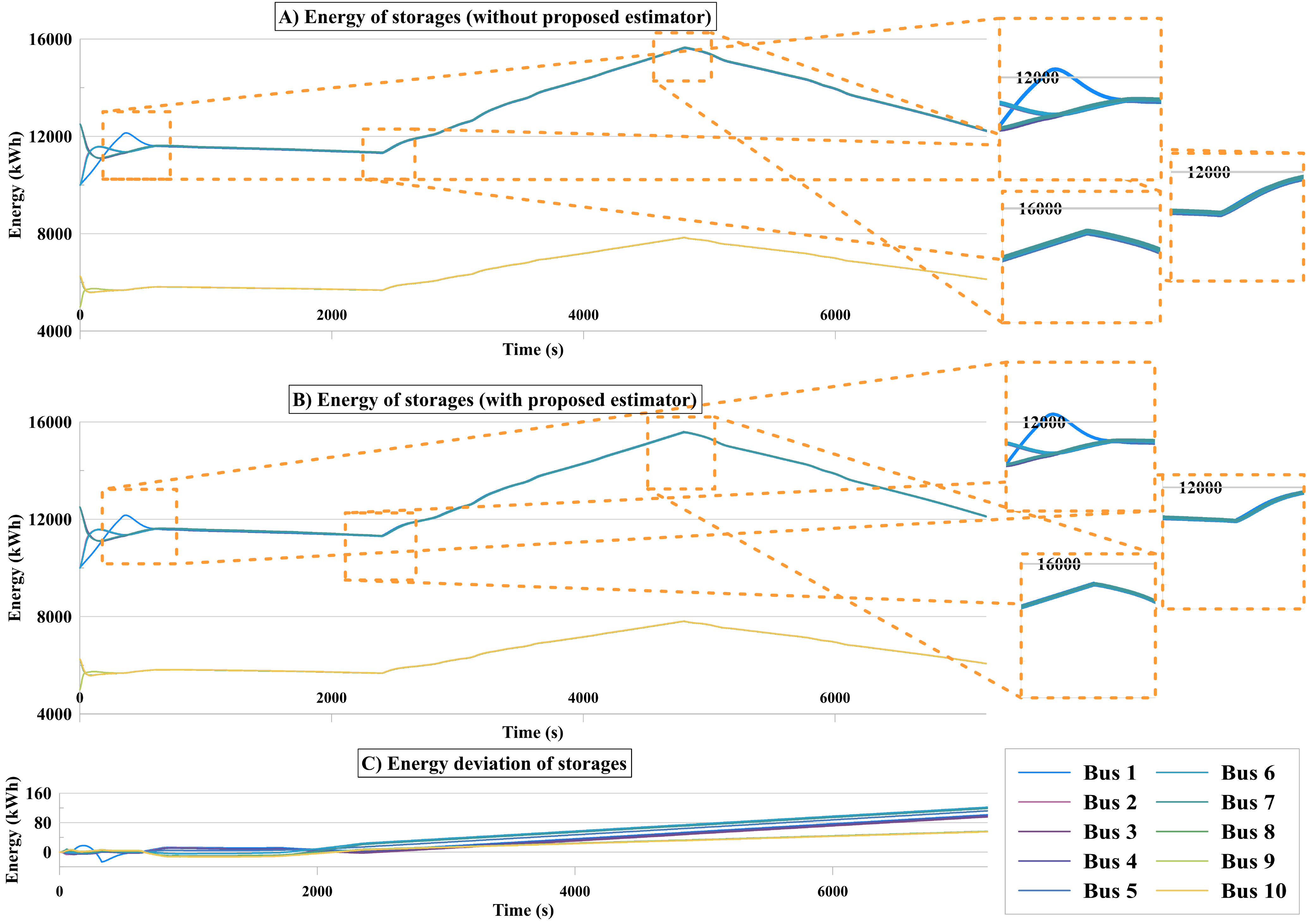}
\caption{Simulation results of the energy level of the storages for the case study of the 10-bus DC microgrid: A) without the proposed estimator; B) with the proposed estimator; C) error between the two approaches.}
\label{simenergy}
\end{figure*}

\begin{figure*}[tb]
\includegraphics[width=14cm,height=11cm]{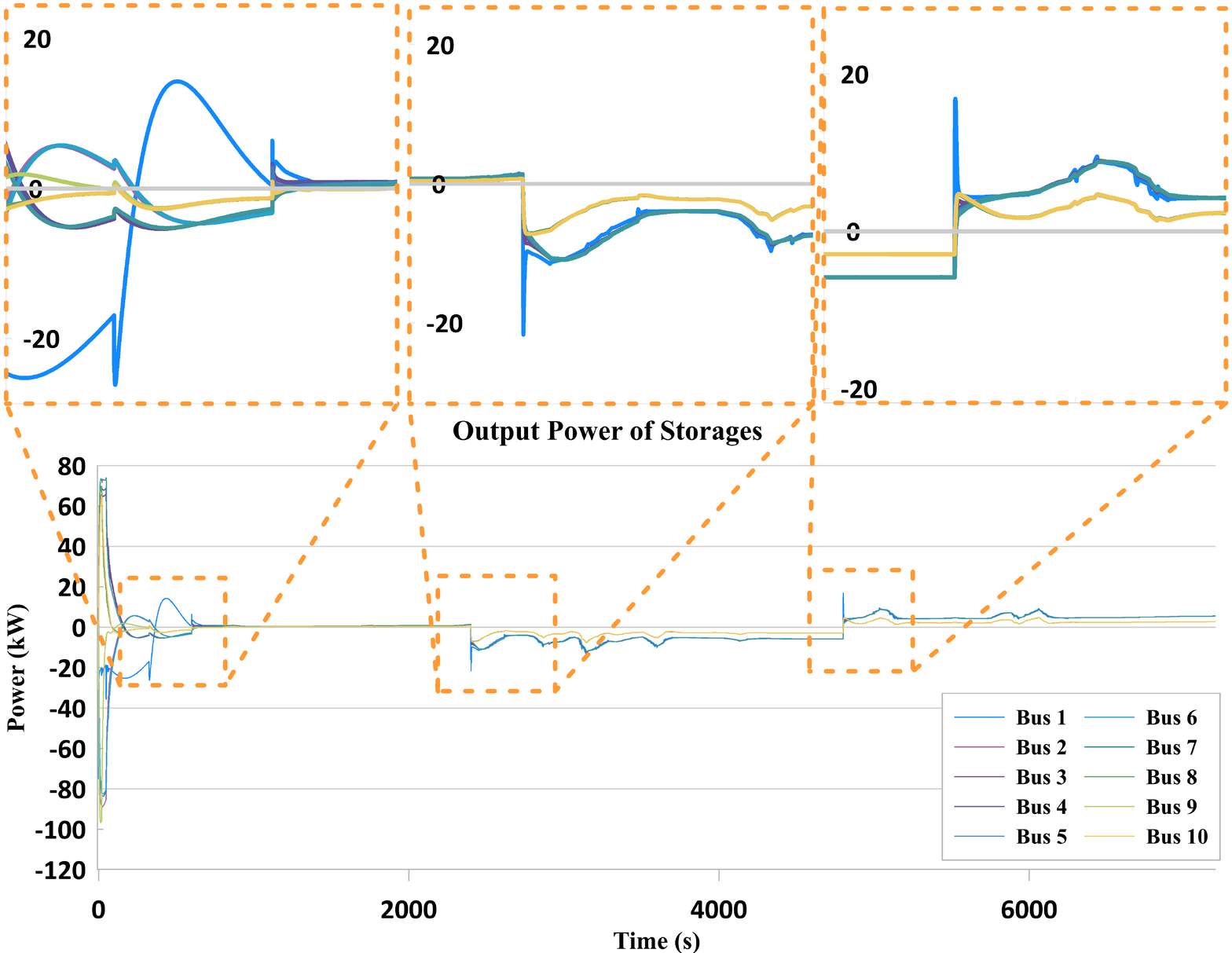}
\centering
\caption{Simulation results of the output power of storages using the proposed distributed event-triggered estimator.}
\label{simpower}
\end{figure*}

Furthermore, by manually adding the delay of 100 ms in the event transmission via the communication network, it can be seen in \figurename ~\ref{simenergy} that the stability throughout the simulation is maintained. It should be noted that the event-triggered control is more prone to the instability, due to the delay in the event-transmission. This is the reason the voltage profile is higher compared with the ideal scenario. This fact is reflected in the error graph of \figurename ~\ref{simenergy} (C).

In the simulation without applying the event-based Kalman filter, each sensor is driven by a synchronous clock, as in a traditional digital control system. For instance, with a sampling period of 1 ms, there should be 40,000 events in the time frame of 40 s. For this specific case study, in 7,200 seconds there should be $7,200 \times 1000 = 72 \times 10^5$ number of generated messages at each bus, but this is much lower with the event-triggered control strategy proposed in this work. The number of generated events at each distributed controller unit of each bus is given in Table \ref{tab:eventscounttbl}. This also shows that the acceptable performance of the microgrid is kept with a minimum number of transactions among the neighboring controllers. The total energy loss in the communication is compared in Table \ref{tab:energycompare}, where the average power consumption of each transceiver is assumed to be 50 mA in the duration of 10 ms at the voltage of 3.3 V (i.e., $e = v \times i \times t$), and the sampling interval is 100 ms for the time-triggered control. When comparing the values of the consumed energy in the nodes in the traditional sample-based control, as well as the proposed event-based one, the effectiveness of the event-based control strategy in terms of the utilization of the resources is evident, as the energy lost in the communication is reduced nearly by 40\%. Additionally, the network traffic is considerably reduced comparing the number of packets generated at each bus with the proposed control strategy to the traditional sample-based control scheme.

\begin{table}[!t]
\renewcommand{\arraystretch}{1.3}
\caption{Generated Events at Each Bus.}
\label{tab:eventscounttbl}
\centering
\begin{tabular}{|c|c|}
\hline
Bus  &  Number of Published Messages \\
\hline
Bus 1 &  44,064 \\
\hline
Bus 2 &  44,159 \\
\hline
Bus 3 &  43,920 \\
\hline
Bus 4 &  43,942 \\
\hline
Bus 5 &  44,198 \\
\hline
Bus 6 &  44,180 \\
\hline
Bus 7 &  44,060 \\
\hline
Bus 8 &  44,184 \\
\hline
Bus 9 &  44,693 \\
\hline
Bus 10 &  43,945 \\
\hline
\end{tabular}
\end{table}

\begin{table}[!t]
\renewcommand{\arraystretch}{1.3}
\caption{Comparison of the Energy Cost Between the Time-Triggered and Event-Based Control Implementations.}
\label{tab:energycompare}
\centering
\begin{tabular}{|c|c|c|}
\hline
Traditional Digital Control  & 720,000 messages total & 0.34 Wh \\
\hline
Event-Based Control &  441,345 messages total & 0.2084 Wh\\
\hline
\end{tabular}
\end{table}

\section{Concluding Remarks}\label{conc}
This investigation has thoroughly presented the design and performance evaluation of a novel distributed event-triggered control and estimation strategy. The objective of this controller is to effectively stabilize the voltage of a DC microgrid only by controlling the output voltages of the DC-DC converters connected to the ES systems. The controller is able to balance the energy level of the ES systems and to regulate the output voltage of the microgrid. An event-based Kalman filter has been developed for the state feedback controller of the DC-DC converters. The Kalman measurement update algorithm has also been modified for the distributed controllers to exchange the data over industrial WSNs. The publish-subscribe model has been proposed for the optimal implementation of the distributed controller, in which the publishers send the data to the specific subscribers without having a subscription knowledge of each node. This has consequently resulted in the intelligent data exchange, as well as the self-configuration of the devices. The simulated results confirm a significant reduction in the network traffic, while maintaining the performance threshold comparing to the digital control schemes. The total energy cost at each sensor node is considerably reduced compared to the traditional time-triggered sampling control systems. This work can be extended to include the comparative analysis of the existing protocols for the proposed controller, along with the performance evaluations using the other estimation methods.

\ifCLASSOPTIONcaptionsoff
  \newpage
\fi

\bibliographystyle{myIEEEtranbibstyle}

\bibliography{IEEEabrv,ref}

\begin{IEEEbiography}[{\includegraphics[width=1in,height=1.25in,clip,keepaspectratio]{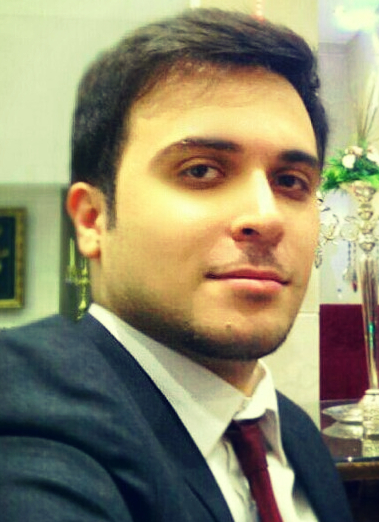}}]{Seyed Amir Alavi}
(GS'16) received the B.Sc. and M.Sc. degrees in electrical engineering from Power and Water University of Technology (PWUT), Tehran, Iran, and Shahid Beheshti University (SBU), Tehran, Iran, in 2013 and 2017, respectively. He is currently pursuing the Ph.D. degree in electronic engineering at Queen Mary University of London (QMUL). He is the supervisor of the Real-Time Microgrid Lab. at QMUL. He has conducted a number of industrial projects on implementation of smart energy management services based on IoT solutions. He is also working as an embedded systems designer in a London-based electricity data intelligence company Voltaware. His research interests include embedded control systems, microgrids, event-based control and signal processing, networked control systems, and wireless sensor networks.
\end{IEEEbiography}

\begin{IEEEbiography}[{\includegraphics[width=1in,height=1.25in,clip,keepaspectratio]{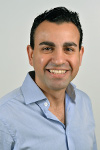}}]{Kamyar Mehran}
is a Lecturer in Power Engineering at Queen Mary University of London (QMUL), U.K. He is the founder of the Real-Time Power and Control System (RPCS) Lab. at QMUL. He is active in the field of intelligent control, fuzzy systems, and multi-sensory fusion systems with a number of publications and book chapters in the field. His research interests include energy storages (vehicular, grid), low-cost DC/AC microgrids, SiC-based inverters, and energy management systems. His recent research grants include a condition monitoring system for aged Lithium-Ion batteries in rural micro-grids (EPSRC Global Challenge Research Fund), and Innovate UK Faraday Battery Challenge “Current Density Imaging in EV battery modules”. He worked as a research fellow in University of Warwick (2013-2015), and Newcastle University (2010-2013). He was the commercialization manager for a spin-off company, OptoNeuro Ltd., commercializing disruptive GaN devices for optoelectronics. Prior to his academic career, he collected over 8 years of industrial experience in companies like Sun Microsystems (Oracle) and National Iranian Oil Company. He received B.Sc. (Tehran, 99), M.Sc. (Newcastle, 2004) and Ph.D. (Newcastle, 2010) all in electronic and electrical engineering.
\end{IEEEbiography}

\vskip -15pt plus -1fil

\begin{IEEEbiography}[{\includegraphics[width=1in,height=1.25in,clip,keepaspectratio]{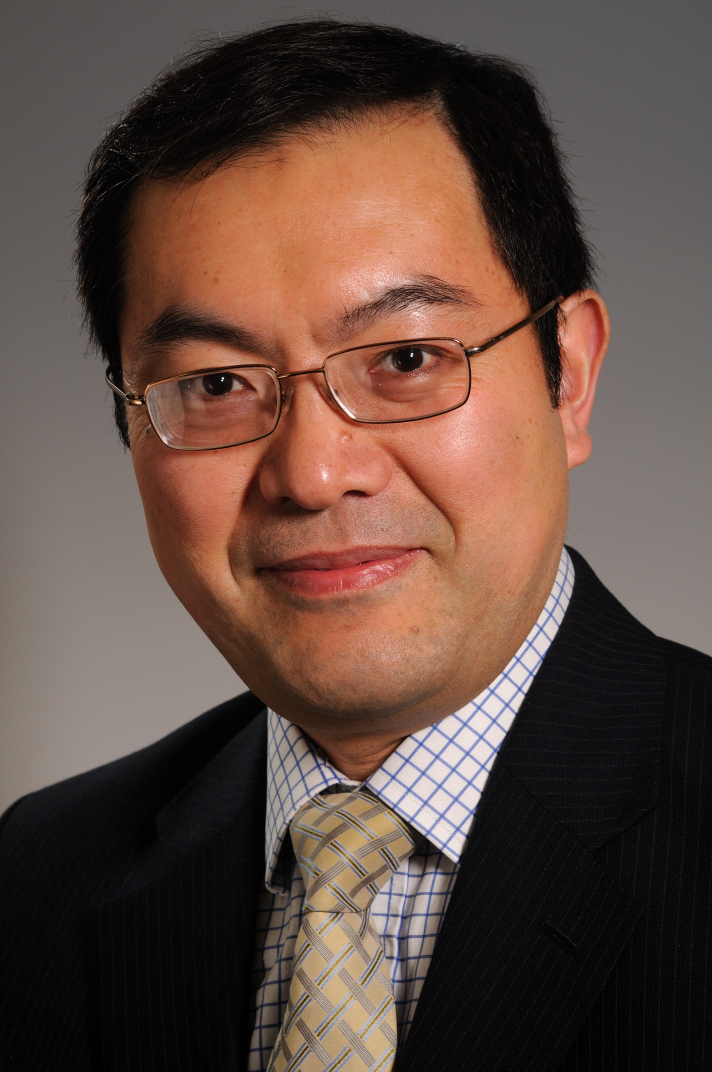}}]{Yang Hao}
(M'00-SM'06-F'13) received the Ph.D. degree on computational electromagnetics from the Centre for Communications Research (CCR) at the University of Bristol, U.K., in 1998.

He is currently a Professor of antennas and electromagnetics in the Antennas \& Electromagnetics Research Group, Queen Mary University of London. Prior to his appointment at QMUL, he was a postdoctoral research fellow at the School of Electronic, Electrical and Computer Engineering, University of Birmingham.

Over the years, Prof. Hao developed several fully-integrated antenna solutions based on novel artificial materials to reduce mutual RF interference, weight, cost, and system complexity for security, aerospace, and healthcare. He developed, with leading UK industries, novel and emergent gradient index materials to reduce mass, footprint, and profile of low frequency and broadband antennas. He co-developed the first stable active non-Foster metamaterial to enhance usability through small antenna size, high directivity, and tunable operational frequency. He has coined the term body-centric wireless communications, i.e., networking among the wearable and implantable wireless sensors on the human body. He was the first to characterize and include the human body as a communication medium between on-body sensors using surface and creeping waves. He contributed to the industrial development of the first wireless sensors for healthcare monitoring. He is a strategic advisory board member for the Engineering and Physical Sciences Research Council (EPSRC), where he is committed to championing RF/microwave engineering for reshaping the future of UK manufacturing and electronics.

Prof. Hao is active in a number of areas, including computational electromagnetics, microwave metamaterials, graphene and nanomicrowaves, antennas and radio propagation for body centric wireless networks, active antennas for millimeter/sub-millimeter applications, and photonic integrated antennas. He has published over 140 journal papers, and he was a co-editor and co-author of the books Antennas and Radio Propagation for Body-Centric Wireless Communications (Boston, MA, USA: Artech House, 2006, 2012), and FDTD Modeling of Metamaterials: Theory and Applications (Boston, MA, USA: Artech House, 2008), respectively.

Prof. Hao is currently a founding Editor-in-Chief of EPJ Applied Metamaterials, a new open access journal devoted to applied metamaterials research. He was an Editor-in-Chief of the IEEE Antennas and Wireless Propagation Letters during 2013-2017. He was an Associate Editor of the same journal, IEEE Transactions on Antennas and Propagation during 2008-2013, and also a Co-Guest Editor for the IEEE Transactions on Antennas and Propagation in 2009. He is a member of Board of the European School of Antenna Excellence, a member of EU VISTA Cost Action and the Virtual Institute for Artificial Electromagnetic Materials and Metamaterials, Metamorphose VI AISBL. He was a Vice Chairman of the Executive Team of IET Antennas and Propagation Professional Network. He has served as an invited and keynote speaker, conference General Chair, session chair and short course organizer at many international conferences. He won the IET A F Harvey Prize in 2016, BAE Chairmans Silver award in 2014, and Royal Society Wolfson Research Merit Award in 2012. He was elected as a Fellow of the ERA Foundation in 2007, Fellow of the IET in 2010, and Fellow of the IEEE in 2013.
\end{IEEEbiography}

\vskip -15pt plus -1fil

\begin{IEEEbiography}[{\includegraphics[width=1in,height=1.25in,clip,keepaspectratio]{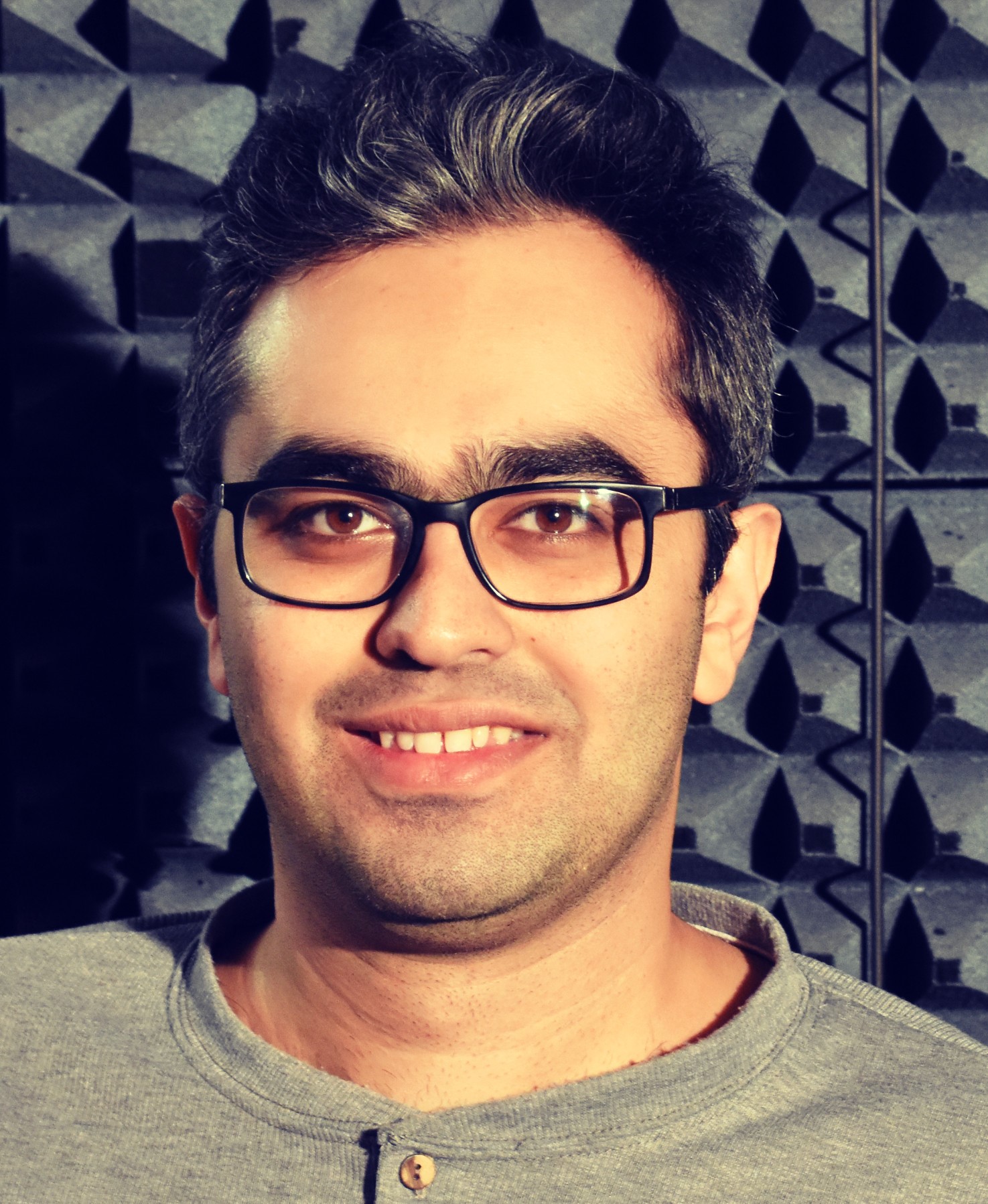}}]{Ardavan Rahimian}
(S'07-M'09) received the M.Eng. (Hons.) degree in electronic and communications engineering from the University of Birmingham, U.K., in 2009. He is currently pursuing the Ph.D. degree in electronic engineering at the Queen Mary University of London, U.K. He was the winner of the Rohde \& Schwarz Technology Prize in 2009. His research interests include antennas, microwaves, millimeter-waves, smart grid, and wireless communications. He has authored or co-authored a number of papers in these areas.
\end{IEEEbiography}

\vskip -15pt plus -1fil

\begin{IEEEbiography}[{\includegraphics[width=1in,height=1.25in,clip,keepaspectratio]{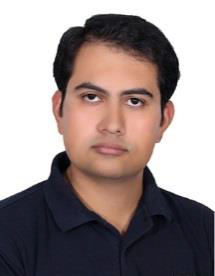}}]{Hamed Mirsaeedi}
(S'16) received the B.Sc. and M.Sc. degrees both in electrical engineering from Power and Water University of Technology (PWUT), Tehran, Iran, and the University of Tehran, Iran, in 2013 and 2017, respectively. He works as a research assistant at the System and Machine Research Lab. (SMRL), Control and Intelligent Processing Center of Excellence (CIPCE), School of Electrical and Computer Engineering, University of Tehran. His research interests include reliability-centered maintenance, asset management, distribution system reliability, smart grid, and power system analysis.
\end{IEEEbiography}

\vskip -15pt plus -1fil

\begin{IEEEbiography}[{\includegraphics[width=1in,height=1.25in,clip,keepaspectratio]{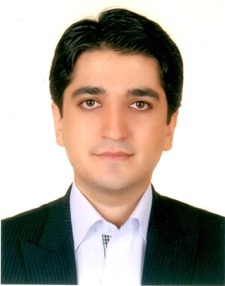}}]{Vahid Vahidinasab}
(M'10-SM'17) received the B.Sc. degree from K. N. Toosi University of Technology (KNTU), Tehran, Iran, in 2004, and the M.Sc. and Ph.D. degrees from Iran University of Science and Technology, Tehran, Iran, in 2006 and 2010, respectively, all in electrical engineering. He is currently an Assistant Professor of electrical engineering at Shahid Beheshti University (SBU). He is the founder and director of the SOHA Smart Energy Systems Lab. at SBU. He has demonstrated a consistent track record of attracting external funds, and he has managed several projects and has closely worked with several large complex national/international projects.

His research interest is oriented to different research and technology aspects of energy systems integration, including the integration of storage and renewable energy resources, PHEVs, demand response, IoT, smart grid/microgrids/nanogrids design, operation and economics, power markets, reliability assessment of power components/systems, and application of artificial intelligence and optimization methods in power system studies.
\end{IEEEbiography}

\end{document}